\documentclass[acmsmall,manuscript,natbib=false]{acmart}
\AtBeginDocument{%
  }

\setcopyright{acmlicensed}
\copyrightyear{2026}
\acmYear{2026}
\acmDOI{XXXXXXX.XXXXXXX}

\acmConference[Conference acronym 'XX]{Make sure to enter the correct
  conference title from your rights confirmation emai}{June 03--05,
  2018}{Woodstock, NY}
\acmISBN{978-1-4503-XXXX-X/18/06}

\RequirePackage[
  datamodel=acmdatamodel,
  style=acmnumeric,
  ]{biblatex}

\usepackage{tablefootnote}
\usepackage{longtable}
\usepackage[table]{xcolor}
\usepackage{tabularx}
\usepackage{enumitem}
\setlist{nolistsep}
\definecolor{green}{HTML}{66FF66}
\definecolor{myGreen}{HTML}{009900}
\usepackage{caption}
\usepackage{subcaption}
\usepackage{xcolor}

\definecolor{red}{named}{black}
\begin{document}

\title{``He gets to be the fun parent'': Understanding and Supporting Burnt-Out Mothers in Online Communities}

\author{Nazanin Sabri}
\email{nsabri@ucsd.edu}
\orcid{0000-0002-0861-9444}
\affiliation{%
  \institution{University of California, San Diego}
  \country{USA}
}

\author{Ananya Malik}
\email{malik.ana@northeastern.edu}
\affiliation{%
  \institution{Northeastern University}
  \country{USA}
}

\author{Bangzhao Shu}
\email{shu.b@northeastern.edu}
\affiliation{%
  \institution{Northeastern University}
  \country{USA}
}

\author{Jason Jeffrey Snyder}
\email{snyder.jas@northeastern.edu}
\affiliation{%
  \institution{Northeastern University}
  \country{USA}
}

\author{Laurie Kramer}
\email{l.kramer@northeastern.edu}
\affiliation{%
  \institution{Northeastern University}
  \country{USA}
}

\author{Mai ElSherief}
\email{m.elsherif@northeastern.edu}
\orcid{0000-0003-4718-5201}
\affiliation{%
  \institution{Northeastern University}
  \country{USA}
}

\renewcommand{\shortauthors}{Sabri et al.}

\begin{abstract}
Maternal burnout is a psychological phenomena with documented harms to both mother and child, requiring prompt attention. Mothers experiencing burnout might choose to turn to online anonymous platforms, such as Reddit, to share their experience, due to feelings of shame and stigmatization of mental health issues. In this work, we study how mothers use Reddit to discuss their experiences of burnout. We first identify posts written by burnt out mothers by manually annotating Reddit posts and training machine learning models on them. Focusing on posts made by this population (N = 3,244), we then investigate the issues brought up by mothers, such as the need for help, career advice, and co-parenting issues. Additionally, we investigate how the Reddit community responds to these posts through the analysis of comments. We find that commenters frequently share personal lived experiences with the poster, and provide emotional support. Finally, considering co-parenting could be a mitigating factor for parental burnout, we explore co-pareting patterns experienced by burnt out mothers, finding evidence of lack of support for and unequal expectations from mothers. 
\end{abstract}



\received{6 February 2026}

\maketitle

\section{Introduction}
\label{section:intro}

Motherhood is a socially constructed identity shaped by cultural norms, gendered expectations, and structural constraints. While many mothers find caregiving rewarding, they also constitute a population facing compounded vulnerabilities, including heightened rates of chronic stress, mental health concerns such as anxiety and depression, and reduced access to formal support systems~\cite{barclay1997becoming, berle2004challenges}. When parents don't have the resources required to manage the stressors that come with having a child, it could lead to burnout~\cite{mikolajczak2020parentalIs}. Parental burnout, defined as ``a condition characterized by intense exhaustion related to parenting, emotional distancing from one's children, and a loss of parental fulfillment"~\cite{mikolajczak2020parentalMove} has become a well-recognized condition in the literature in recent years~\cite{mikolajczak2021beyond}. This type of burnout can have pervasive negative effects on the individuals' experiences of being parents, as well as their interactions with their children~\cite{mikolajczak2020parentalMove}. Parental burnout can impact both mothers and fathers. However, in most societies, parenting is still largely a gendered-role with mothers being expected to take on more of the parenting duties~\cite{bastiaansen2021gender}. As a result, in this work, we focus on maternal burnout (i.e., burnout experienced by mothers)\footnote{We acknowledge the existence and importance of paternal burnout and the need for further research in that area. Nevertheless, maternal health is a critical topic in need of attention as it is ``a perfect storm of healthcare vulnerabilities, with historical biases and power dynamics influencing care"~\cite{antoniak2024nlp}.}. 

\begin{figure}
    \centering
    \includegraphics[width=\linewidth]{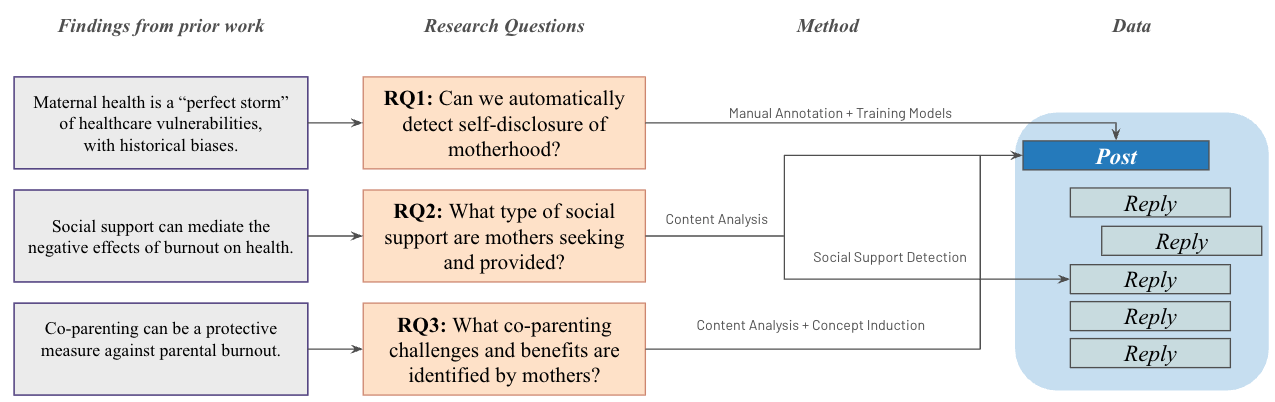}
    \caption{The motivations, research questions, methods, and data used in this work.}
    \label{fig:research_questions}
\end{figure}

While some mothers turn to loved-ones or in-person groups for support, dominant societal narratives that often idealize motherhood\footnote{Pressures to be a ``perfect mother", for instance, have been positively associated with burnout~\cite{meeussen2018feeling}.} can lead to stigma around emotional disclosure and barriers to seeking help~\cite{hubert2018parental, kirkpatrick2022comparisons, henderson2016price}. Maternal burnout may contribute to mothers' reluctance to participate in studies, underscoring the need for systematic, scalable, and minimally intrusive approaches to understanding their perspectives and emotional experiences. In recent years, social media platforms have become a pervasive part of daily life, offering spaces where individuals can connect, exchange support, and share personal stories. Design affordances of these platforms encourage the self-disclosure of emotionally sensitive experiences, namely: anonymity, asynchronous communication, and selective audience visibility~\cite{ammari2019self}. This is especially valuable in contexts where expressing vulnerability may carry social stigma or challenge prevailing cultural ideals about motherhood. Mothers increasingly turn to social media to voice their struggles, seek advice, and find community with others facing similar challenges~\cite{porter2013mothers}. These platforms serve as informal support networks, offering validation, empathy, and shared understanding—particularly for those who may lack access to traditional mental health resources or who feel isolated in their caregiving roles. 


A quick search on the internet reveals large number of groups and websites that are being used by mothers who might be struggling (e.g., mumsnet, what to expect). Reddit is one of the websites that is used by burnt out mothers to vent out their frustrations, and seek advice. For instance, popular subreddits such as ``r/Parenting" (\# members = 7.5M), ``r/breakingmom" (\# members = 127K), and ``r/workingmoms" (\# members = 110K) are dedicated to the discussion of the ups and downs of parenting. 
Prior work has investigated online postings of parents on such forums~\cite{10.1145/2675133.2675149, ammari2019self, 10.1145/2531602.2531603}. However, majority of the existing work have focused on the general use and content posted by all mothers, and not specifically those dealing with issues such as burnout.

This study aims to address the challenges of accessing mothers experiencing burnout 
at scale by examining public online parenting communities on Reddit, through the analysis of 16 years of self-initiated, open-ended conversations. Through a theory-driven lens, prior work on maternal health vulnerabilities, the protective role of social support, and the benefits of co-parenting inform our research questions~\cite{antoniak2024nlp, velando2020impact, bastiaansen2021gender}. First, building on the understanding that maternal health is shaped by historical biases and systemic neglect~\cite{antoniak2024nlp}, we examine the phenomenon of self-disclosure in online motherhood narratives, addressing the current gap in NLP methods for automatically detecting such emotionally and contextually rich expressions. Then, recognizing the effect of social support on maternal stress and burnout~\cite{velando2020impact}, we examine the types of support mothers seek and receive within these communities. Lastly, informed by research on co-parenting as a protective factor~\cite{bastiaansen2021gender}, we explore how mothers describe the challenges and benefits of co-parenting in their narratives (Figure \ref{fig:research_questions} displays these motivations, research questions, and methods of our work). We summarize our three research questions below:

\begin{itemize}
    \item[\textbf{RQ1}] Can we automatically detect self-disclosure of motherhood?
    \item[\textbf{RQ2}] What types of social support are mothers seeking and provided?
    \item[\textbf{RQ3}] What co-parenting challenges and benefits do posters describe?
\end{itemize}

In order to answer \textit{RQ1} we annotated a subset of 1,800 posts (from a 28K dataset of posts discussing burnout) for authorship by a mother. Among other considerations, to avoid misidentification, we carefully designed a codebook to capture all forms of ``motherhood". Using the annotated data, we created automated models to detect motherhood at scale. Our best model achieves a macro F1 score of 0.91. We discuss the implications of these models in the Discussion Section~\ref{discussion:more_models}. Using our best trained model, we labeled the entire dataset to indicate authorship by mothers. A subset of 3,244 posts were predicted to fit our criteria of authorship by burned out mothers. These data points are then used to answer the remainder of our research questions. We present the results of this study in Section \ref{section:study1:method}.

Next, to answer \textit{RQ2} we performed a mixed-methods analysis of the 3,244 posted by burnt out mothers. Through a manual annotation of 100 randomly sampled posts, we identified 4 distinct needs being asked of the online communities these mothers post to: advice, being heard, celebrating success stories, and diagnosis. To analyze the support being provided by the community, we analyzed the comments these posts receive. We discuss the topics discussed within these comments, as well as classifying them based on the informational or emotional support they provide. We found that emotionally supportive responses are more common. Commenters are also likely to share their own personal struggles, potentially with a hope of helping the poster feel less alone. These results are presented in Section \ref{section:study2}.

Finally, to answer \textit{RQ3}, we performed manual annotations and trained automated models to extract all portions of posts that discussed a posters' co-parent. A total of 3,366 sentences were extracted through this step. Next, we performed topic modeling and qualitative analysis on these sub-sections of posts to better understand the co-parenting issues mothers bring up. We found unequal gender-based expectations, weaponized incompetence, and miscoordination between co-parents to be some of the issues burnt out mothers face. These results are presented in Section \ref{section:study3}.


In summary, this work examines how maternal burnout is discussed in online spaces, revealing key challenges mothers face and the support they seek. Our findings contribute to research on maternal burnout, online mental health discourse, and the role of digital communities. We conclude our work by offering recommendations to inform the design of future tools for supportive, context-aware interventions (Section \ref{section:discussion}).

\section{Related Work}
\label{section:related_work}
In this section, we will situate our work within two areas of related work. First, in Section \ref{section:related_work:parental_burnout}, we provide a brief summary of prior studies of maternal burnout. Next, we situate our work within existing studies investigating discussions of mental health in online spaces in Section \ref{section:related_work:online_discourse}. 

\subsection{Parental Burnout}
\label{section:related_work:parental_burnout}

Burnout is defined as a ``prolonged response to chronic emotional and interpersonal stressors on the job, and is defined by the three dimensions of exhaustion, cynicism, and inefficacy"~\cite{maslach2001job}. This response was first identified in the context of occupations, specifically those working in health and social service~\cite{maslach2006burnout}. Recent studies, however, have identified burnout in other populations and contexts such as parental burnout~\cite{mikolajczak2019parental}. Additionally, studies have suggested that burnout is context-related, with context playing a meaningful role in the consequences of being burnt out~\cite{mikolajczak2020parentalIs}. 

Parental burnout is ``characterized by an overwhelming exhaustion related to one's parental role, an emotional distancing from one's children, and a sense of parental ineffectiveness"~\cite{mikolajczak2019parental}. This phenomena is distinct from ordinary parental stress~\cite{mikolajczak2020parentalMove}. Parental burnout has been shown to strictly increase escape ideation, as well as neglectful and violent behaviors toward one's children~\cite{mikolajczak2019parental}. Initial studies of parental burnout focused on parents of children with illnesses or disabilities~\cite{bilgin2009reducing, lindahl2007burnout}. However, more recent studies have broadened the scope of parental burnout, looking at all parents. In 2017, the proportion of parents experiencing burnout was reported to be between 2\% and 12\%~\cite{roskam2017exhausted}. These rates have been shown to vary based on country, with cultural values playing an important role~\cite{roskam2021parental}. Another study, for instance, found maternal burnout rates to be 20\%~\cite{sejourne2018maternal}. Notably, with regard to the differences, individualistic cultures were shown to have higher prevalence of parental burnout~\cite{roskam2021parental}. Studies have also looked at the factors that can cause mothers to feel burnt out. For instance, prior work has found fear (e.g., of not being a good enough mother) to be a central factor in the experience of burnout by mothers~\cite{hubert2018parental}. 
Factors that could help improve burnout have also been studied. For instance, co-parenting has been reported as a mitigating factor~\cite{bastiaansen2021gender}. 

Scholars in feminist theory and maternal studies have long critiqued how cultural narratives around motherhood obscure the structural pressures and emotional labor that mothers endure. The ideology of intensive mothering, for instance, constructs motherhood as a selfless, all-consuming identity, setting unrealistic expectations that can lead to guilt, isolation, and emotional exhaustion~\cite{ennis2014intensive}. Critical motherhood studies challenge these norms and advocate for centering the lived experiences of mothers themselves~\cite{mack2018critical}. 


\textcolor{red}{\textbf{Our Contribution.} Our work builds on studies of maternal burnout, performing a content analysis of burnt out mother's self-disclosure on online communities (Reddit). The majority of prior work has used surveys and interviews for the study of this phenomena. Given the stigma around the topic of maternal burnout, the study of voluntary self-disclosures on pseudo anonymous platforms could allow us to discover needs mothers might otherwise not be comfortable sharing. Additionally, the use of this data might provide us access to insights from sub-populations of mothers who might be unable to take part in studies (e.g., too overwhelmed, or not in the regional focus of study).}

\subsection{Online Disclosure of Mental Health}
\label{section:related_work:online_discourse}

The rise of social media platforms and online peer counseling forums has revolutionized the methods by which people share knowledge and communicate. These online communities allow individuals to seek support anonymously from trained or untrained support providers on a variety of topics, including mental health~\cite{Naslund:2016, milton2023see, chancellor2019human, pendse2023marginalization, boettcher2021studies, 10.1145/3710903, pendse2025role, 10.1145/3710904}, parental experiences~\cite{lee2025designing, 10.1145/2470654.2470700, 10.1145/2702123.2702205, sabri2025posting}, and women's needs~\cite{10.1145/3710978}. With this in mind, researchers have leveraged social media data to computationally predict mental health status and improve mental health outcomes of at-risk individuals~\cite{Chancellor:2020}. A large body of work has investigated mental health~\cite{10.1145/3584931.3607013, 10.1145/3678884.3682051, 10.1145/3678884.3681871, 10.1145/2818048.2819973} on a variety of platforms, including Twitter~\cite{berry2017whywetweetmh, mcclellan2017using}, Facebook~\cite{egan2013college}, Reddit~\cite{de2014mental}, and many more~\cite{johnsen2002online, feuston2018beyond, 10.1145/3711089}. 

Reddit, with 116 million daily active users as of 2025\footnote{\url{https://www.demandsage.com/reddit-statistics/}}, is among the popular platforms for receiving support. Looking at three Reddit mental health communities, Park et. al. found four shared themes of discussion, namely: (1) sharing of positive emotion, (2) gratitude for receiving emotional support, (3) sleep- and (4) work-related issues~\cite{park2018examining}. Fraga et. al. found that interactions across subreddits are similar and centered around content rather than users~\cite{fraga2018online}. Studies have also looked into the impact of the COVID-19 pandemic on discussions on mental health subreddits~\cite{biester2021understanding, low2020natural}. They report that while there is no notable increase in number of submissions on these subreddits, the content being shared has changed considerably~\cite{biester2021understanding} (e.g., health anxiety becoming an emerging shared theme within Reddit submissions after the pandemic~\cite{low2020natural}). Prior work has also studied TikTok videos that included diagnostic steps \& treatments~\cite{10.1145/3584931.3607013}. 

Topically related to our work is Sabri et. al.'s study of expressions of burnout on Reddit~\cite{sabri-etal-2024-inferring}. The authors developed a dataset of burnout expressions and models to detect the self-disclosure of burnout given text-based input~\cite{sabri-etal-2024-inferring}. It is important to note that this study was not focused on experiences of burnout by a specific community (e.g., mothers). Additionally, the authors were more focused on the development of a tool for identification of these posts rather than an analysis of the discourse. Other related works include \cite{Saha:2017}, \cite{saha2019language}, \cite{10.1145/2531602.2531675}, and \cite{Cascalheira_Hamdi_Scheer_Saha_Boubrahimi_Choudhury_2022} which assess the self-disclosure of stress to develop a greater understanding of stigmatized topics within online discourse. More specifically, \cite{saha2019language} and \cite{Cascalheira_Hamdi_Scheer_Saha_Boubrahimi_Choudhury_2022} ground themselves in Minority Stress Theory~\cite{frost2023minority} to better explore the nuanced language used to express stressors within the experiences of individuals identifying as LGBTQ+~\cite{saha2019language}. Meanwhile, Saha et. al. examined expressions of stress from survivors of gun violence on college campuses within online discourse~\cite{Saha:2017}. Finally, using Facebook data, as well as self-reports and validated postpartum depression status, De Choudhury et. al. developed models to predict the onset of postpartum depression~\cite{10.1145/2531602.2531675}.

Prior work has also investigated use of technologies for mental health by a variety of populations such as veterans~\cite{10.1145/3584931.3607009}, those struggling with eating disorders~\cite{10.1145/2818048.2819973}, international students~\cite{10.1145/3678884.3681871}, and mothers~\cite{10.1145/3678884.3682051, sabri2025posting, gergely2015mothers, teaford2019new, 10.1145/3311957.3359481}. Mothers have been shown to turn to a number of social media platforms to share different aspects of their lives~\cite{gergely2015mothers, sabri2025posting}. Studying Facebook groups, Gergely found that this space was used by mothers to (a) ask questions, (b) promote events, (c) share photos, and (d) draw attention, among other things~\cite{gergely2015mothers}. Mothers of young children were also shown to ask questions about parenting purchases, social interactions, social support, and favors on Facebook~\cite{morris2014social}. Teaford et. al. investigated postings during the first 6 months postpartum, identifying five themes: (1) social support, (2) anonymity, (3) in-groups, (4) drama, and (5) entertainment/pastime~\cite{teaford2019new}. Virtual mothering, an emerging way through which mothers perform online, was identified with four aspects of: perfection, privacy, politics, and play~\cite{bailey2023virtual}. Among studies that focus on the use of online spaces, cluster analysis of profiles of mothers have also been conducted, identifying groups experiencing greater difficulties~\cite{lebert2021and, sanchez2019depression}. Mothers' engagement in 3 WhatsApp groups made up of pregnant women, new mothers, and medical practitioners were investigated in~\cite{10.1145/3311957.3359481}. The authors found that participants used these groups for both casual chats, and asking health-related questions, finding that posters expressed frustration when responses to their questions were not prompt~\cite{10.1145/3311957.3359481}. The analysis of the content of these groups also revealed a preference for expert rather than peer knowledge~\cite{10.1145/3311957.3359481}. Interviews with medical practitioners who were part of these groups also showcased an exhaustion trying to keep up with expectations, respond to repeated questions, and separate questions from chatter~\cite{10.1145/3311957.3359481}. 

Prior studies have also looked into how platform affordances and design might impact what is shared on each one. For instance, how mothers of children younger than three used Facebook and Twitter was investigated by Morris (2014), through survey responses and data dumps by 412 participants~\cite{10.1145/2531602.2531603}. The authors found that Twitter was not viewed as an appropriate medium to share information about one's child, however, mothers engaged in indirect discussion by sharing links to news articles or stories relevant to their child~\cite{10.1145/2531602.2531603}. Facebook, on the other hand, was used to share updates and images of participant's children, with the goal of questioning, curation, and ``mommy networking"~\cite{10.1145/2531602.2531603}. The authors found that the questions asked on Facebook were with the aim of receiving advice, social support, parenting purchases, social interactions (e.g., asking if people wanted to join them for different activities), and asking for favors~\cite{10.1145/2531602.2531603}. 
Motivations of sharing baby-content online was studied by~\cite{10.1145/2675133.2675149}. The authors found that while risks of sharing content were acknowledged and mentioned by the mothers they interviewed (e.g., oversharing, managing the child's digital footprint), the benefits of this sharing (e.g., archiving childhood, receiving validation, and identifying as a mother) outweighed their concerns~\cite{10.1145/2675133.2675149}. Zhou et. al. studied mom vloggers, looking into motivations for becoming vloggers and challenges faced~\cite{10.1145/3584931.3606977}. The authors reported prior work experience, making money, and creation of community as reasons for becoming a vlogger as a mother~\cite{10.1145/3584931.3606977}. The interviewed mothers listed getting more views, personal attacks, under- or over-regulation by platforms, and privacy concerns as challenges they faced~\cite{10.1145/3584931.3606977}. 
 

\textbf{Our Contribution.} We build upon and extend work looking into the design of technology for parenting and child development~\cite{10.1145/3719160.3737631, 10.1145/3022198.3026334, 10.1145/3022198.3024940}. Our work extends earlier studies by focusing specifically on maternal burnout which is an under-explored topic in computational social science, and bridges research on mental health~\cite{10.1145/3710904, 10.1145/3710903, 10.1145/3173574.3174215}, motherhood~\cite{zhu2024maternal, meeussen2018feeling, teaford2019new}, and social support~\cite{10.1145/3512980, 10.1145/3025453.3026008, 10.1145/3411764.3445203}. Additionally, our work continues the investigation of platform affordances, focusing on Reddit, specifically in the use case of mothers experiencing burnout. We explore how Reddit's affordances such as communities focused on specific issues, throwaway accounts, and engagement with posts through comments, impact these users. Through mixed-method analysis we investigate the mothers' expressed goals and needs for posting, as well as the types of responses and support they receive.


\section{Study 1: Detecting Self-disclosure of Motherhood}
\label{section:study1}
Maternal health and well-being is an issues recognized in the scientific literature~\cite{souza2024global, filippi2006maternal} and by policy makers (e.g., United Nations sustainable development goals\footnote{https://sdgs.un.org/goals/goal3}) alike. Burnout is one of the issues that this population could be affected by~\cite{lebert2018maternal}. In this work, we aim to provide a better understanding of the challenges and issues faced by burnt out mothers. To be able to understand these challenges, we first needed to be able to detect self-declarations of motherhood and burnout at scale. Thus, in our first study we investigate how mothers express their identities through text, and create automated models for the detection of these expressions. 

\subsection{Method}
\label{section:study1:method}

\subsubsection{Data Collection \& Cleaning}
\label{section:study1:method:data_collection}

As we were interested in the characterization of discourse on subreddits, we began by collecting Reddit submissions spanning the years 2008 to 2023\footnote{Collected from existing archives of Reddit data, made available for research purposes.}. Given our focus on the intersection of burnout and motherhood, we restricted our dataset to subreddits dedicated to discussions around \textit{mental health} and \textit{parenting}.
Prior work has introduced lists of subreddits dedicated to mental health and parenting~\cite{10.1145/3173574.3174215, 10.1145/3411764.3445203}. We expanded these initial lists by conducting targeted searches on Reddit, allowing us to compile a comprehensive set of subreddits relevant to our focus. A total of 41 mental health and 30 parenting subreddits were considered, 22 of which do not appear in prior work ~\cite{10.1145/3173574.3174215, 10.1145/3411764.3445203}. The full list of the subreddits used in this work is provided in Table \ref{table:list_of_subreddits}. After retrieving posts from these subreddits, we removed submissions with bodies shorter than 10 characters, resulting in a dataset of 4,565,162 posts. 

\begin{table*}[h]
    \centering
    \begin{tabular}{p{1.2cm}|p{13cm}}
        \textbf{Category} & \textbf{List of Subreddits} \\\hline\hline 
         Parenting (N = 30) & r/breakingmom, r/beyondthebump, r/daddit, r/raisingKids, r/toddlers, r/stepparents, r/parentingteenagers, r/lgbtfamilies, r/cisparenttranskid, r/regretfulparents, r/workingmoms, r/breastfeeding, r/breastfeedingsupport, r/pottytraining, r/clothdiaps, r/downsyndrome, r/homeschool, r/predaddit, r/radicalparenting, r/atheistparents, r/parentsofmultiples, r/boobsandbottles, r/parent, r/teachingchildren, r/birthcontrol, r/parentdeals, r/playgroup, r/thingsmykidsaid, r/cutekids, r/vbac\\\hline 
         Mental Health (N = 41) & r/mentalhealth, r/depression, r/adhd\_anxiety, r/socialanxiety, r/anxiety\_support, r/malementalhealth, r/depression\_help, r/depressionregimens, r/getting\_over\_it, r/ptsd, r/schizophrenia, r/autism, r/lonely, r/addiction, r/Bingeeating, r/Phobias, r/schizoaffective, r/selfharm, r/mentalillness, r/alcoholism, r/traumatoolbox, r/dbtselfhelp, r/therapy, r/burnedout, r/abuse, r/adultsurvivors, r/afterthesilence, r/bullying, r/domesticviolence, r/emotionalabuse, r/rapecounseling, r/survivorsofabuse, r/dpdr, r/psychoticreddit, r/calmhands, r/helpmecope, r/hardshipmates, r/offmychest, r/reasonstolive, r/depressed, r/itgetsbetter\\\hline 
    \end{tabular}
    \caption{List of subreddits dedicated to mental health or parenting that were used to filter the data used in this work. Some subreddits are no longer visible on Reddit, but are included in the archive of posts.}
    \label{table:list_of_subreddits}
\end{table*}

Having created this dataset, we conducted a second round of filtering by retaining only submissions that contained at least one burnout-related keyword in either the title or body\footnote{We decided to identify posts discussing mental burnout using keyword searches, rather than machine learning models, because the domain of our posts are restricted to mental health and parental subreddits. While the word burnout can be used to refer to electrical burnout or the \href{https://en.wikipedia.org/wiki/Burnout_(series)}{game series burnout}, we believe these use cases would be much less prominent in the domain of our data.}. We explore different stemming functions of two Python text analysis libraries (NLTK and Spacy) to find a function that would convert different versions of ``burn out" to its basic form. Some examples of these different shapes include: \textit{burnout, burn out, burnt out, burned out}, and \textit{burning out}, with considerations of the use of space or `-` between `burn' and `out'. To ensure that all versions of burnout are correctly mapped to their base version, we randomly sampled 100 posts and investigated how different libraries changed the mention of this phrase. Our final pipeline, using Spacy~\cite{spacy2}, ensured that all variations would be taken into account within our keyword-search.  

By focusing on posts that explicitly mention burnout, we aim to center self-expressions and lived experiences, rather than making assumptions or attributing mental health concerns without individuals' own articulation. Users could still be mistaken about the assignment of burnout to themselves, especially considering the potential confusion between postpartum depression and burnout~\cite{confusion_between_burnout_and_depression}. However, we argue that these self-expressions would be more valid compared to assigning burnout to users who have not self-disclosed. Self-disclosure as ``the telling of the previously unknown so that it becomes shared knowledge"~\cite{10.1145/2702613.2732733}, is recognized as an important step for improvement~\cite{10.1145/2702613.2732733}, and has been the focus of a large number of studies by the HCI community~\cite{joinson2007self, kou2018you}. This filtering reduced the number of data points to analyze to 28,346 submissions (0.6\% of the initial set).

Table \ref{table:d1_stats} displays the basic statistics of the filtered dataset of 28K posts. As seen in the table, the majority of posts are written by unique authors, with the overall average number of posts per author being 1.24 (this average was calculated when `deleted' and `AutoModerator' were considered within the mix). 2,519 posts were made by users who had deleted their account after posting. 58 posts had also been posted by \textit{AutoModerator}\footnote{Posts made by AutoModerator are often listing the rules of the subreddit. One example of a post by AutoModerator is \href{https://www.reddit.com/r/breakingmom/comments/1kcpt25/breakingmom_rules_reminder/}{available here}. Additional information about what AutoModerator is and how it works can be \href{https://www.reddit.com/r/reddit.com/wiki/automoderator/}{found here}.}. Having identified these authors, we remove posts made by \textit{AutoModerator}. This removal of posts by \textit{AutoModerator} reduced the number of posts in parental subreddits to 4,192. The number of posts in mental health subreddits did not change. We chose to keep posts by users who had deleted their account, because many users use \textit{throwaway accounts} to discuss sensitive topics such as mental health issues (e.g., burnout)~\cite{ammari2019self}. Additional information about this dataset is provided in Appendix \ref{appendix:methods:study1:subreddit_list}. 

\subsubsection{Ethical Considerations for Researching Public Online Data}
\label{section:study1:method:ethical_considerations}

Throughout this work we use publicly accessible Reddit posts. A large number of studies have used public online data for the analysis and characterization of various phenomena~\cite{abbar2015you, kucuktunc2012large, magdy2015failedrevolutions}, including topics related to mental health~\cite{biester2021understanding} and motherhood~\cite{sanchez2019depression} (we provide an overview of these studies in our related work presented in Section \ref{section:related_work:online_discourse}). While we use this data to provide a better understanding of the challenges faced by mothers who had disclosed experiencing burnout, we do not assign the labels of ``mother" or ``burnout" to posters if there is no self-declaration by the poster. We only use the data as-is and do not make additional contact with the posters. 
We acknowledge the complexities of online and automated maternal health research, as explored by prior work~\cite{10.1145/3653690, vitak2017ethics, antoniak2024nlp}. We believe this work is attempting to give a voice to those seeking help, thus aligning with the the guidelines stated by prior work~\cite{antoniak2024nlp}. 

\begin{table*}
    \centering
    \begin{tabular}{l c c c c c | c c}
         &  &  &   & \multicolumn{2}{c}{\textbf{Title}} & \multicolumn{2}{c}{\textbf{Body}} \\\cline{5-8}
        \textbf{Category} & \textbf{\# Posts} & \textbf{\# Users} & \textbf{\# Subreddits}  & \textbf{Char.} & \textbf{Words} & \textbf{Char.} & \textbf{Words}\\\hline 
         Parenting & 4,250 & 3,464 & 21 & 41 ($m.=35$) & 9 ($m.=8$) & 1,783 ($m.=1,391$) & 399 ($m.=312$) \\
         Mental Health & 24,096 & 19,361 & 35 & 48 ($m.=40$) & 10 ($m.=9$) & 2,000 ($m.=1,368$) & 440 ($m.=304$)  \\\hline 
    \end{tabular}
    \caption{Average and median (denoted by $m.$) character and word lengths for post titles and bodies in our dataset. The number of posts reflects the number of Reddit submissions after filtering for posts that had burnout keywords (in their title or body) and had bodies (i.e., ``selftext'') with more than 10 characters. The burnout keywords used for filtering are different variations of \textit{burnout} identified through text processing and stemming. Title and body length values have been rounded to the closest decimal value.}
    \label{table:d1_stats}
\end{table*}

\subsubsection{Qualitative Annotation of Motherhood}
\label{section:study1:method:qualitative_annotation}

To analyze mothers' experience of burnout, we needed to find the subset of posts written by mothers. However, preliminary exploration of the use of large language models (LLMs) to detect authorship by mothers revealed that these models used stereotypes and biases (e.g., having a child or domestic labor) to assign motherhood. As a result, we decided to create a manually labeled dataset in which posts are annotated for explicit signs of self-disclosure of motherhood. More concretely, we annotated a post for being written by a mother if one of the following conditions is true: (1) the author explicitly says that they are a mother (e.g., ``I am a mother", ``being a mother, I"), or (2) the author mentions being a woman (e.g., ``as a woman", ``32F") \textbf{and} having children (e.g., ``my kid"). Implicit signs such as breastfeeding and being pregnant are considered as signs of being a mother. However, whether or not the poster gave birth was not a consideration. As long as the poster self-identified as a mother, they were labeled as a mother. For instance, step-parents are included in our analysis. Similarly, when it comes to womanhood, the sex assigned at birth to the poster was not considered and instead we rely on the gender the poster identifies as and declares in their post. Finally, the use of phrases like ``my husband" is not taken as signs of motherhood as no assumptions about the sexuality of the individual are made. 

To create our annotated dataset, we first sampled 1,200 posts at random from the dataset of 28K posts identified above(Table~\ref{table:d1_stats}). 600 submissions were sampled from mental health subreddits and 600 from parental subreddits. This sampling created $Batch1$ of our labels (Table~\ref{table:annotation_counts}). These posts were then annotated by 3 members of our research team. The annotation team was made up of one graduate computer science student (with prior experience in computational social science and interdisciplinary projects) and two undergraduate computer science students. Prior to annotation, the three students met and agreed on the methods and definitions of the annotation task. Annotators were able to see the subreddit name, title, and body of a post when making decisions about their label. The three annotators began by all labeling 30 posts from batch 1 for being authored by a mother or not (binary labels). Annotators had 100\% agreement on these thirty posts\footnote{The labels assigned by all three annotators matched each other for all 30 posts.}. As a result, after a short meeting in which annotators came together and reviewed the details of the task, they proceeded to label the remainder of the posts individually. All annotators labeled roughly an equal number of Reddit posts (around 400 posts each). Through the annotation of $Batch1$, we found that the majority of the subset of posts that were written by mothers were posted to parenting subreddits (with only a small number coming from mental health subreddits). As a result, we then randomly sampled and labeled an additional set of 600 submissions, this time sampling all posts from parenting subreddits ($Batch2$). Table~\ref{table:annotation_counts} shows the number of posts by mothers in each batch. In accordance with our sampling strategy, $Batch2$ includes more positive (1) labels. 

\begin{table}
    \centering
    \begin{tabular}{l|c|c||c}
        \textbf{Label} & \textbf{Batch 1} & \textbf{Batch 2} & \textbf{Total}\\\hline
         Mother (1) & 316 (26\%) & 302 (50\%) & 618 (34\%)\\
         Not a Mother (0) & 884 & 298 & 1,182\\\hline\hline
         Total & 1,200 & 600 & 1,800 \\
    \end{tabular}
    \caption{Number of posts within each batch of manually annotated data for authorship by a mother.}
    \label{table:annotation_counts}
\end{table}

\subsubsection{Detection of Authorship by a Mother}
\label{section:study1:method:automated models}

To detect motherhood at scale (i.e., on the 28K posts), we trained a number of deep learning and in-context learning models, which have been recognized as potentially useful models for computational social science tasks~\cite{ziems2024can, wang2024instruction}. We decided against using regular expressions, as our observation when annotating posts was that disclosures of motherhood did not follow strict patterns that would be easily captured using regular expressions. We will discuss the details of training our ML models and hypterparameter tuning in this section. We used \textit{sklearn}'s \textit{train\_test\_split} function to split our dataset of 1,800 annotated posts into train, test, and validation sets. We first split the data into train and test sets using a split of 20\% using stratified sampling. We then used stratified sampling on the training data with a split of 10\% to create the validation set. These two steps resulted in three sets with the following label distributions (0 = not a mother, 1 = a mother):

\begin{itemize}
    \item Training data (72\% of total): (\# 0 = 851, \# 1 = 445)
    \item Validation data (8\% of total): (\# 0 = 95, \# 1 = 49)
    \item Test data (20\% of total): (\# 0 = 236, \# 1 = 124)
\end{itemize}

The same three sets of data are used for all models. The concatenation of post title and body are provided as input for all models. The performances reported in the body of the work are on the test set. However, all hyperparameter tuning efforts were preformed on the validation set. Once the best set of hyperparameters were discovered, the model was trained on the entire training set (including the validation data) and then deployed on the test set. We used \textit{sklearn}'s \textit{classification\_report} and \textit{confusion\_matrix} functions to evaluate the performance of our models. 


\noindent\underline{\textit{3.1.5.1 Deep Learning Models:}} We fine-tuned a number of bert-based models using the combination of hyperparameters listed in Appendix \ref{appendix:methods:study1:dl_hyperparams}. The best performing model on the validation set was \textit{microsoft/deberta-v3-base} with the following hyperparameter values: $max\_length: 512$, $learning\_rate: 2e^{-05}$, $num\_train\_epochs: 4$, $per\_device\_train\_batch\_size: 16$, $weight\_decay: 0.01$. This model achieved an F1 score of 0.97 on the validation set. 

\noindent\underline{\textit{3.1.5.2 In-Context Learning Models:}} We tested zero-shot and few-shot strategies in our work using \textit{GPT-4o} as this model has been shown to have good performance on computational social science tasks~\cite{ziems2024can}. Our prompt (displayed in Table~\ref{tab:zero-shot-prompt}) reflects the diverse nature of motherhood and specifically calls out different types of mothers present in the literature, such as stepmothers~\cite{cann2018very}, pregnant mothers~\cite{rahmawati2019influence}, surrogate mothers~\cite{jacksonmother} and breastfeeding/nursing mothers~\cite{marshall2007being}. Our zero-shot learning prompt is shown in Table~\ref{tab:zero-shot-prompt}. We do not use chain-of-thought prompting as it is less effective in computational social science tasks~\cite{shaikh2022second}. All models were tested with $temperature = 0$, as we are not looking for creative outputs. Using a temperature of zero for computational social science tasks to ensure receiving consistent and reproducible results is recommended by prior work~\cite{ziems2024can}. 

\begin{table}
    \centering
    \begin{tabular}{@{}p{\textwidth}@{}}
    \hline 
    \emph{Zero-shot prompt} \\\hline
    \textbf{Title}: ``Classification of authorship of text"  \\
    \textbf{Definition}: In this task, we ask you to classify the input text into two options:\\
(A): Written by a mother: the poster includes details about themself that indicate they are a mother, such as explicitly saying they are a mother or discussing their gender as a woman and having children.\\
(B): Not written by a mother: the poster does not explicitly indicate being a mother.\\
    \textbf{Emphasis} $\&$ \textbf{Caution}: Only look for explicit signs indicating the poster is a mother, stereotypes or roles should not be used to indicate motherhood and should NOT be labeled as (A). Furthermore, a stepmom, a pregnant mother, a surrogate mother and a breastfeeding/nursing mother can also be classified as a mother. \\
    \textbf{Things to avoid}: All input must be classified into one of the options. If you cannot pick then choose the option with higher probability. The output must be either (A) or (B) but not both.\\
    \textbf{Input}: $\{$text$\}$ \\
    \textbf{Output}: \\
    \hline
    \end{tabular}
    \caption{Zero-shot prompt for detection of authorship by a mother.}
    \label{tab:zero-shot-prompt}
\end{table}

We used a similar prompt for few-shot learning. We tested $number\_of\_exemplars \in \{2, 4, 6, 8\}$ to identify the optimal number of examples that allows the model to achieve its best performance. We used two strategies to select examples for our few-shot learning approach:

\begin{itemize}
    \item \textbf{Random}: We select $n$ exemplars at random from the training set. The random selection is performed for each data point individually. The only rule we enforce on the random selection is that $\frac{n}{2}$ of examplars need to have $label = 1$ and the other half need to have $label = 0$.
    \item \textbf{KNN}: We first use the OpenAI embeddings model \textit{text-embedding-3-small} on the query data point and all data points in the training set. Next, we use KNN~\cite{mucherino2009k} with $k= number\_of\_exemplars$ to find the closest data points within the training set that more closely resemble our data point and provide those to the model. It is worth noting that an equal number of exemplars for each label is not enforced in this method. We use sklearn's implementation of KNN (\textit{KNeighborsClassifier}) for this step. 
\end{itemize}

\subsection{Findings}

In this section, we report the results of the models trained on our annotated dataset. Table \ref{table:best_performing_models} displays the performance of different models on the test set. As seen in the table, a fine-tuned version of \textit{deberta-v3-base} that is trained on our training set, outperforms large language models like \textit{GPT-4o} in zero-shot and few-shot settings. We perform error analysis for our best in-context learning and deep learning models to better understand their limitations. This error analysis is presented in Appendix \ref{appendix:methods:study:error_analysis}.   


\begin{table*}
    \centering
    \begin{tabular}{l c|c c c c| l}
        \textbf{Model} & \textbf{\# Exemplars}& \textbf{Accuracy} & \textbf{Precision} & \textbf{Recall} & \textbf{F1} & \textbf{Distribution of Test Set}\\\hline\hline
        Zero-shot & - & 0.75 & 0.79 & 0.81 & 0.75 & \# mother (1) = 124\\
         &  & &  &  &  &  \# not mother (0) = 236\\\cline{1-6}
        Few-shot (Random) & 2 & \textbf{0.86} & \textbf{0.85} & \textbf{0.88} & \textbf{0.86} &  \\
         & 4 & 0.84 & 0.85 & 0.89 & 0.84 &  \\
         & 6 & 0.83 & 0.83 & 0.87 & 0.83 &  \\
         & 8 & 0.81 & 0.81 & 0.85 & 0.80 & \\\cline{1-6}
        Few-shot (KNN) & 2 & 0.81 & 0.82 & 0.85 & 0.81 &  \\
         & 4 & 0.75 & 0.79 & 0.81 & 0.75 &  \\
         & 6 & 0.74 & 0.78 & 0.80 & 0.74 &  \\
         & 8 & 0.75 & 0.79 & 0.81  & 0.75 & \\\hline\hline 
        deberta-v3-base & - & \textbf{0.92} & \textbf{0.91} & \textbf{0.91} & \textbf{0.91} & \\\hline 
    \end{tabular}
    \caption{The performance of our deep learning and in-context learning models on the test set. Macro F1 scores are being reported. The bold rows represent the highest performing deep learning and in-context learning models based on F1 scores.}
    \label{table:best_performing_models}
\end{table*}

Given the high performance of the \textit{deberta-v3-base} model, and considering this model has less cost (compared to the GPT model), we train this model on the entire annotated dataset. Once the model was trained, we deployed it on the collected 28K posts to predict whether each post was written by a mother or not. Our final dataset is the subset of posts predicted to be written by mothers ($prediction = 1$). However, we decided to include all posts written in ``r/breakingmom" subreddit in the set authored by mothers (regardless of the label assigned by the deep learning model) as this subreddit has a rule stating that ``only mothers" can post in it\footnote{We checked the rules of the most common subreddits we encountered and this subreddit was the only one with such a rule. Additionally, a manual inspection of a subset of posts within this subreddit confirmed that this rule does seem to be enforced (i.e., we did not observe posts by individuals who are not mothers).}. Overall, a total of 3,244 (11.4\%) Reddit submissions were labeled as authored by mothers. These submissions consisted of 2,704 posts from subreddits dedicated to discussion of parental needs and 540 from subreddits dedicated to the discussion of mental health issues. 
We use this dataset (N = 3,244) in our next two studies.

\subsubsection{Model Generalizability Testing} To test the extent to which the performance of our motherhood-detection model is able be generalized we performed one additional test. In this test, we investigated the performance of the model on posts that do not have mentions of burnout. The goal of this step was to see if the performance of our motherhood-detection model is dependent on the context of the post being burnout-related. Accordingly, we constructed a sample of 100 Reddit posts, drawn from an out of sample dataset, that do not contain any lexical variants of the term burnout. These 100 posts were sampled from all Reddit submissions posted in December 2023. 50 of these posts were from subreddits that have ``parent" as part of their name, while another 50 were randomly sampled from other subreddits. We made sure that subreddits with ``mom" in their name are not included in either set. The decision to exclude these subreddits was made to make this set more distinct from the domain of the training data\footnote{Thus providing greater insight into the true generalizability of our model.}. We first performed a manual annotation of these 100 posts. Among the created sample, 7 were identified to be written by mothers, and the other 93 to not be written by mothers. However, it is worth nothing that among the 93 that are labeled as \textit{not written by a mother}, a number of them discussed family issues and issues with the poster's mother (however, the posters themselves did not disclose identifying as mothers). 

Having created this semi-random sample of 100 posts, we used our trained \textit{deberta-v3-base} model to predict the authorship of these posts. The performance of this model is as follows: $Accuracy = 0.94$, $Precision = 0.77$, $Recall = 0.84$, $F1 = 0.8$. We did not test the performance of our best in-context learning model as we did not deploy that model on the dataset. We can see that while performance decreases slightly, the model still performs acceptably on out-of-context posts. A manual inspection of the errors made by the model reveals that posts in which the author refers to their mother without disclosing their own motherhood are among the ones most likely to be misclassified. However, it is important to note that this misclassification does not always happen. We believe future work should explore more datasets and models for the detection of motherhood within text in order to better address maternal needs. We further discuss the implications of our work and future directions in Section \ref{section:discussion}. 

Finally, this analysis supports the validity of our decision to annotate for motherhood disclosure within a pre-filtered subset of posts that explicitly mention burnout. By first narrowing our dataset to posts containing burnout keywords before conducting manual annotations for motherhood, we ensured that our model was trained within the context most relevant to our research goals. Since our primary aim was to better understand the experiences of mothers experiencing burnout, it was more critical for the model to perform well in this specific domain rather than on general posts. Notably, our model achieves a strong in-domain macro F1 score of 0.91, compared to an out-of-domain F1 score of 0.80. While the latter indicates promising generalizability, reversing the annotation order—labeling for motherhood prior to filtering for burnout—might have reduced performance on the domain that matters most to our analysis.

\section{Study 2: Exploring Self-Disclosure of Maternal Burnout and Social Support Dynamics}
\label{section:study2}
The goal of this work is to learn about the experiences of burnt out mothers, and create a better understanding of their wants and needs. This investigation will provide online community organizers, moderators, and health professionals with a better understanding of the challenges faced by burnt out mothers, allowing them to provide better support to this group~\cite{gore2024maybe}. Thus, having identified posts written by mothers who self-disclosed burnout, in our second study, we explore how they express themselves and the needs they have from the online community they are talking to. Additionally, we analyze the responses (i.e., comments) burnt out mothers receive, investigating the types of support being provided by other members of these online communities.  

\subsection{Method}

\subsubsection{Understanding Mothers' Needs} To explore the needs expressed in burnt out mothers' posts, we performed a mixed-method analysis: (i) qualitative and (ii) quantitative. Using this mixed-method approach allows us to form a deep understanding of the issues being discussed (using i) while also discovering overarching and repeated patterns across our large dataset (using ii). Methodologically, our content analysis approach is similar to prior work~\cite{10.1145/3653690, 10.1145/3512980}.

For our qualitative analysis (i), we randomly sampled 100 posts from all posts classified as being written by mothers and using burnout keywords (N = 3,244). These posts were sampled at random without any restrictions. Once these 100 posts were selected, one member of our research team (a graduate computer science student with prior experience in qualitative analysis) performed a manual analysis of these posts, performing inductive coding~\cite{chandra2019inductive}. For each post, the annotator was provided the id, title, and body of the post.
The annotation focused on extracting what the poster was directly or indirectly asking of their readers. It is important to note that the annotator was labeling for needs from the online community, rather than from people in the poster's life. For instance, segments such as ``I need my partner to do X" are not captured, while segments such as ``I need advice on how to ask my partner to do X" are included in the coding. This is because we are interested in the needs the poster has of the online community, trying to better understand why they come to this space and choose to post online. A total of 8 categories were extracted from these posts which were then grouped into 4 needs (Table \ref{table:needs_qualitative}). 

To be able to analyze all posts (N = 3,244), and understand the overall content of posts, we also performed a quantitative content analysis of the posts (ii). To do so, we performed topic modeling, using BERTopic~\cite{grootendorst2022bertopic}. This method has been used by a large number of studies to discover discussion patterns in social media data~\cite{bhuvaneswari2024topic, dan2025exploring, hswen2024experiences, mobin2024exploratory, ng2024hype}. We explored different hyperparameters for \textit{min\_samples} and \textit{min\_cluster\_size} and select the model with the highest coherence score. We experimented with performing topic modeling on the body or the title of posts. Our findings revealed more coherent topics when only titles were used. We hypothesize that the relatively poor performance observed when using the full post body as the unit of analysis may be attributed to the long length of the body and the presence of multiple, potentially unrelated topics within a single post, which may dilute the signal relevant to the task. Additionally, as posters often state their main question or goal for posting in the title, using titles is appropriate for our goal. We also tested LLooM~\cite{lam2024conceptInduction}, an LLM-based concept induction algorithm meant to extract high-level concepts from unstructured data, on our dataset. However, we found the topics produced by LLooM to be too generic and high-level (e.g., generating ``burnout'' or ``emotional exhaustion'' as topics). 

\subsubsection{Collecting Comments} We
collected all comments associated with the 3,244 posts written by burnt out mothers. 239 (7\%) posts did not have any comments\footnote{Collected from existing archives of Reddit data, made available for research purposes.}. For the remaining 3,005 posts, a total of 50,674 comments (written by 21,816 unique users) were collected. Among these comments, 8,575 (16\%) were responses written by authors of the posts themselves (i.e., the poster had commented on their own post responding to someone else or providing additional context). As we are interested in the support provided to posters, we decide to focus on the first-level comments (i.e., comments posted as direct replies to the original post, rather than those in reply to other comments). This filtering resulted in 27,044 first-level comments (53\%). While we do offer some statistics for the entire set of comments (see Table \ref{tab:comment_support_type}), the majority of our work focuses on the subset of first-level comments. 

\subsubsection{Understanding the Content of Comments} We performed a similar two-step (i) qualitative and (ii) quantitative analysis of the comments to understand their content and the type of support they provide. Our qualitative analysis was done through inductive coding on a random sample of 100 first-level comments. Our quantitative analysis was done using a fine-tuned BERTopic model on all first-level comments. 

\subsubsection{Automatic Detection of Support in Comments} To further understand the type of support provided to mothers, we trained informational and emotional support detection models. These types of social support have received attention in the social computing literature~\cite{10.1145/3512980, 10.1145/3025453.3026008}. 
Informational support ``is the information, advice, referrals or knowledge that people exchange"~\cite{10.1145/3025453.3026008}, while emotional support is defined as ``the expressions of caring, concern, reassurance, or empathy people receive from others"~\cite{10.1145/3025453.3026008}. To train models capable of assessing the presence of emotional support (ES) or informational support (IS) in a comment, we used labeled Reddit comments from a previous study on Reddit comments~\cite{10.1145/3173574.3174215}. The dataset includes 397 comments, randomly sampled from 55 mental health subreddits labeled 1 (least supportive of the original poster) to 3 (most supportive) for informational and emotional support~\cite{10.1145/3173574.3174215}\footnote{Within this dataset, each comment has two labels: One for emotional and one for informational each a number between 1 and 3.}.  

As we were most interested in evaluating whether a comment is supportive or not, rather than evaluating a comment's level of support, we decided to modify the existing dataset. This change entailed converting the three-class labels for informational and emotional support into binary labels indicating existence (1) or lack (0) of each type of support. Our transformation procedure is explained in Appendix \ref{appendix:methods:study2:social_support_converting}. We fine-tuned two sets of BERT-based models to assess if a comment could be emotionally or informationally supportive. Our hyperparameter tuning and error analysis of our best models for each support category are provided in Appendix \ref{appendix:methods:study2:social_support_models}. The performance and hyperparameters of the best models for each support type is displayed in Table \ref{table:support_model_performance}. We can see that the best model for both classes reaches an F1 score of above 0.8. These models were deployed on all collected comments (N = 50,674). 

\begin{table}[h]
    \begin{tabular}{p{2cm}|c| c c c |c c c c}
        \textbf{Support Type} & \textbf{Model} & \textbf{Learning Rate} & \textbf{Batch Size} & \textbf{Epochs} & \textbf{Accuracy} & \textbf{Precision} & \textbf{Recall} & \textbf{F1}\\\hline
        Emotional & DeBerta & $4e^{-5}$ & $4$ & $4$ & $0.83$ & $0.87$ & $0.87$ & $0.87$\\
        Informational & DeBerta & $4e^{-5}$ & $4$ & $6$ & $0.90$ & $0.85$ & $0.95$ & $0.90$\\
    \end{tabular}
    \caption{Performance of our best support detection models.}
    \label{table:support_model_performance}
\end{table}




\subsection{Findings}
\label{section:study3:findings}

We present the results of this study in three phases: we first explore the content of posts and the needs being expressed. We then discuss the content of comments, and the support being provided to posters. Finally, we provide a brief analysis of the current dynamics of seeking and providing support in these Reddit communities. 

\subsubsection{What are mothers seeking by posting on Reddit?} To answer this question we begin by going over the results of our qualitative analysis, displayed in Table \ref{table:needs_qualitative}. We use the notation of \textit{(S+\textit{number})} to refer to the labeled post we are quoting\footnote{$S$ stands for Sample.}. The number is not associated with the post's ID but is instead the index of the post within the random sample of 100 we examined. A single post can be labeled as expressing multiple needs (thus, the totals in the table exceed 100). We present our findings in this section, and discuss the implications and connections to prior work in Section~\ref{discussion:online_self_disclosure}. 

We found that venting, ranting, or sharing frustrations with the goal of ``getting things off one's chest" was the most common reason posters expressed (49). It appears that mothers might be framing their posts in ways that could potentially encourage more supportive responses.  Asking for advice regarding specific aspects of one's life, such as relationships (N = 19), career (N = 13), and child-care (N = 20) were also common within the posts. Advice seeking is a recognized pattern in Reddit communities~\cite{carpenter2018advice, reagle2025history, adelina2023stories}. 



When posts were made with the goal of venting and sharing one's experience, the posters often expressed not having anyone else they felt like they were able to talk to. For instance, saying ``I don't have anyone I can talk to about this, so I have a lot to get out" \textbf{(S89)}. One poster expressed feeling like they were unable to share their struggles with others because ``I'm so scared of losing my kids" \textbf{(S77)}. Thus the lack of sharing could be both due to potential real consequences, as well as not wanting to appear incompetent or incapable to people who know them personally. At times the frustration being expresses was due to receiving unsolicited advice (e.g., ``I'm so tired of everyone telling me to put my kid in daycare." \textbf{(S69)}). Some posters disclosed feeling unappreciated because no-one was checking in on them or asking about their well-being (e.g., ``And. No. One. Asks. How. I'm. Doing." \textbf{(S95)}). Overall, wanting to be heard by people who have experienced or are going through the same thing as the poster seemed to be a big motivator for posters. As one mother put it:  

\begin{quote}
    I think I just want a space where I can write down how I feel ... To let it out and maybe hear if others have gone through something similar.  \textbf{(S97)}
\end{quote}

During our qualitative analysis, we also observed that mothers frequently include rich contextual information about their living circumstances at the time of posting. These details often encompass the number and ages of their children, the employment status and working hours of both themselves and their partner, the division of household responsibilities, and their emotional perceptions of their partner or co-parenting relationship. Offering this level of detail might be done with the hope of getting more specific and usable advice. We also observe that they often use very strongly negative language towards themselves, for instance: ``I'm a bitter burnt out mother" \textbf{(S23)}. Some posters also displayed cognitive distortions such as all-or-nothing thinking: ``I'm so sick of failing at everything" \textbf{(S19)}. The existence of these instances of negative self-talk are inline with prior work, as a correlation between negative thoughts and burnout has been shown~\cite{falahati2020prediction, chang2017examining}. Having described their struggles and the issues they were facing, some mothers also felt the need to defend themselves, using language such as ``I'm not a terrible mom or wife" \textbf{(S20)}. This pre-emptive defense might be due to a perceived need to manage external judgment or internalized guilt, reflecting the societal pressures and normative expectations surrounding idealized motherhood roles~\cite{hada2021ruddit, 10.1145/3543507.3583522}.

\begin{center}
\begin{table}
\caption{Different types of needs being expressed in the posts made by burnt out mothers based on our qualitative analysis of 100 randomly sampled posts.}\label{table:needs_qualitative}
\resizebox{0.98\columnwidth}{!}{
\begin{tabularx}{\textwidth}[t]{XX}
\arrayrulecolor{green}\hline
\textbf{\textcolor{myGreen}{Need 1: Advice or Opinions}} & \\
\hline
1.A Career Advice: advice about their career such as changing where they work, if their work conditions are sustainable, whether they should accept a promotion, and issues faced after coming back from maternity leave ($N = 13$)& 
\begin{minipage}[t]{\linewidth}%
\begin{itemize}
\item[1.1] ``Do you all think 15 hours a week without childcare, with this setup, is sustainable? Or is this going to lead to burnout?" \textbf{(S4)}
\item[1.2] ``Anyone else returned from mat leave and feeling like they hate working?" \textbf{(S13)}
\item[1.3] ``I am back to work after 4 months of maternity leave and I am so lost in my role." \textbf{(S24)}
\item[1.4] ``Business owners, working moms? [...] Any advice/opinions from current business owners or anyone would be greatly appreciated." \textbf{(S35)}
\end{itemize} 
\end{minipage}\\

\arrayrulecolor{black}\hline

1.B Child Management Advice: advice about how to better manage one's child, such as advice on pumping, childcare, and bonding with one's child ($N = 20$)&
\begin{minipage}[t]{\linewidth}%
\begin{itemize}
\item[1.5] ``Moms of multiple small kids - how do you split childcare on the weekends so you both get a break?" \textbf{(S20)}
\item[1.6] ``I lost it on my 5 year old twice in one week" \textbf{(S17)}
\end{itemize} 
\end{minipage}\\

\hline

1.C Self-Care Advice: seeking recommendations about how to take care of themselves and/or recover from burnout ($N = 8$)&
\begin{minipage}[t]{\linewidth}%
\begin{itemize}
\item[1.7] ``I'm just wondering how [stay at home parents] cope with burnout when there’s literally no way of having a break?" \textbf{(S28)}
\item[1.8] ``How are you fitting in time to workout?" \textbf{(S29)}
\end{itemize}
\end{minipage}\\

\hline

1.D Relationship Advice: asking for advice about dynamics with their family, friends, and partner, include advice on relationship issues or issues with how chores are split ($N = 19$)&
\begin{minipage}[t]{\linewidth}%
\begin{itemize}
\item[1.9] ``Do I need to leave my entire situation? I don't feel valued or appreciated by my partner. I have told him what I need from him but I am still reminding him months later" \textbf{(S5)}
\item[1.10] ``Am I in a one sided friendship? [...] Open to advice, opinions or relatable stories." \textbf{(S10)}
\item[1.11] ``How do I make my husband understand? How do I make him understand I can't go on like this." \textbf{(S64)}
\end{itemize}
\end{minipage}\\

\arrayrulecolor{green}\hline
\textbf{\textcolor{myGreen}{Need 2: Being Heard}} \\
\hline

2.A Venting/Ranting: sharing the frustrations they are experiencing for the sake of sharing ($N = 49$)&
\begin{minipage}[t]{\linewidth}%
\begin{itemize}
\item[2.1] ``I just needed to get this off my chest." \textbf{(S1)}
\item[2.2] ``maybe just needed to rant or find people that commiserate with me" \textbf{(S13)}
\item [2.3] ``thanks for ``listening". I don't really know what I'm looking for in posting this." \textbf{(S72)}  
\end{itemize}
\end{minipage}\\\arrayrulecolor{black}\hline

2.B Seeking Assurance: asking for assurance that how they are feeling is ``normal" ($N = 8$)& 

\begin{minipage}[t]{\linewidth}%
\begin{itemize}
\item[2.4] ``Are these just the hellish years that are hard? Is this normal?" \textbf{(S1)} 
\item[2.5] ``Is this normal?" \textbf{(S6)}
\item[2.6] ``Please tell me I will survive this. Please encourage me and give me anything to keep my mind positive." \textbf{(S100)}
\end{itemize}
\end{minipage}\\

\arrayrulecolor{green}\hline
\multicolumn{2}{l}{%
\textbf{\textcolor{myGreen}{Need 3: Sharing Success Stories}}} \\
\hline

Sharing about how things have improved, potentially in order to give others hope ($N = 3$) &
\begin{minipage}[t]{\linewidth}%
\begin{itemize}
\item[3.1] ``If anyone is on the fence for a job change for the new year, just do it [...] mental health is way more important." \textbf{(S98)}
\end{itemize} 
\end{minipage}\\

\arrayrulecolor{green}\hline
\textbf{\textcolor{myGreen}{Need 4: Seeking Diagnosis}} \\
\hline

Asking for a mental health diagnosis given how they are feelings (i.e., the symptoms they describe) ($N = 2$) &  
\begin{minipage}[t]{\linewidth}%
\begin{itemize}
\item[4.1] ``burnt out, mom guilt, postnatal depression or overreacting?" \textbf{(S1)} 
\end{itemize}
\end{minipage}\\
\arrayrulecolor{black}\hline

\end{tabularx}
}
\end{table}
\end{center}

To further understand the content of posts, and the issues mothers discuss we preformed topic modeling, the results of which are displayed in Table \ref{tab:topic_modeling_on_posts_s3}. We found a total of 15 topics from these posts. The most frequently discussed topics included: venting about burnout $(T_1)$, emotional struggles $(T_3)$, and parenting challenges $(T_{8}, T_{10}, T_{14})$. The high presence of these topics confirms our qualitative findings. Issues related to career decisions $(T_4)$, relationships $(T_5)$, sleep deprivation $(T_6)$, and breastfeeding needs $(T_7)$ were also brought up. Mothers might be asking these questions on Reddit instead of their social circles to make sure that the responses they get are from a population that has itself struggled with similar issues, rather than people who might not have first-hand experience and might offer generic answers. The wide variety of topics also indicates that burnout is not limited to very specific issues experienced by mothers with regards to child-care and is instead all-encompassing, in terms of the areas of life impacted. This finding is inline with existing theories of parental burnout~\cite{mikolajczak2018theoretical, ren2024systematic}. These theories suggest that the factors contributing to parental burnout span a wide range of life domains~\cite{ren2024systematic}. Prior work that investigated parental help-seeking\footnote{The referenced study investigates help-seeking among all parents and not necessarily mothers self-disclosing burnout.} on Mumsnet also found posters discussing parents' emotions and sleep which are among our discovered topics as well~\cite{10.1145/3653690}. 

\rowcolors{2}{gray!25}{white}
\begin{table}[]
    \centering
    \begin{tabular}{c p{6cm} p{7cm}}
        \textbf{Frequency} & \textbf{Topic} & \textbf{Top Keywords} \\\hline    
        941 & Needing Advice or to Vent $(T_1)$ & need, rant, tired, advice, vent, break, feel\\
        \multicolumn{3}{p{15.3cm}}{"Any advice appreciated - LONG rant" ($P_{6}$), "Just a bad day vent" ($P_{162}$)}\\
        739 & Stating Their Burnout $(T_2)$ & burnt, burned, burn, feeling, ppd, burning, beyond, tip, else, feel\\
        \multicolumn{3}{p{15.3cm}}{"I'm burnt out." ($P_{134}$), "At what point is parenting burnout a mental health crisis?" ($P_{1164}$))}\\
        504 & Stating Emotional Struggles $(T_3)$ & burnout, dealing, depression, extreme, hitting, emotion, recover, avoid\\
        \multicolumn{3}{p{15.3cm}}{"Endlessly exhausted. Parental burnout or depression?" ($P_{753}$)}\\
         273 & Discussion of Career $(T_4)$ & job, work, leave, career, time, maternity, quit, back, end, new, breadwinner\\
         \multicolumn{3}{p{15.3cm}}{"Went back to work after maternity leave :'(" ($P_{31}$), "Help me decide my next career move!" ($P_{1966}$)}\\
         197 & Discussion of Various Relationships $(T_5)$ & husband, marriage, friend, relationship, partner, divorce, like, job, spouse, made\\
         \multicolumn{3}{p{15.3cm}}{"15 year friendship over?" ($P_{744}$), "How do I make my husband understand?" ($P_{284}$)}\\
         128 & Discussions of Sleep \& Nursing $(T_6)$ & month, sleep, old, nursing, night, bedtime, bed, baby, every, help\\
         \multicolumn{3}{p{15.3cm}}{"My daughter spends hours awake at night crying." ($P_{2748}$), "When did you give up “the bedtime routine”?" ($P_{2010}$)}\\
         90 &  Discussion of Breastfeeding $(T_7)$ & breastfeeding, pumping, breast, supply, feeding, milk, pump, stop, bottle, baby\\
         \multicolumn{3}{p{15.3cm}}{"Breastfeeding older toddlers" ($P_{1704}$), "Breast pump recommendations" ($P_{494}$)}\\
         73 & Toddler Issues $(T_8)$ & toddler, baby, newborn, daughter, toy, handling, second, covid, leaving\\
         \multicolumn{3}{p{15.3cm}}{"Everything with my toddler is a fight" ($P_{429}$), "I called my toddler a nightmare" ($P_{2046}$)}\\
         71 & Feeling at one's wits end $(T_{9})$ & break, broken, breakdown, end, rope, wit, wound, insufferable, handcuff\\
         \multicolumn{3}{p{15.3cm}}{"I am at the end of my rope with my insufferable toddler" ($P_{1223}$), "I'm edging closer to a breakdown" ($P_{3173}$)}\\
         57 & Challenges \& Experiences as a Stay at home mother (SAHM) $(T_{10})$ & sahm, sahms, burnout, work, back, working, sahp, frustration, long, anyone\\
         \multicolumn{3}{p{15.3cm}}{"What are SAHM responsible?" ($P_{164}$), "I don’t want to be a SAHM anymore, and I feel guilty AF." ($P_{2341}$)}\\
         37 & Discussions of Autism \& ADHD $(T_{11})$ & autistic, autism, adhd, parent, resource, appointment, help, raising\\
         \multicolumn{3}{p{15.3cm}}{"Psychiatrist appointment soon.... Adhd or autism or both?" ($P_{2886}$), "Working mom with autism/ADHD. I'm burning out I think." ($P_{2608}$)}\\
         36 & Discussions of Special Occasions $(T_{12})$ & birthday, christmas, year, holiday, worst, merry, gift\\
         \multicolumn{3}{p{15.3cm}}{"Effing holidays" ($P_{2251}$), "Last minute toddler birthday ideas?" ($P_{962}$)}\\
         35 & Discussions of Pregnancy $(T_{13})$ & pregnant, postpartum, pregnancy, depression, abortion, birth, severe, business, stress, health\\
         \multicolumn{3}{p{15.3cm}}{"Burnout at 9 months pregnant with a 1 year old : advice needed." ($P_{1071}$), "Tired of being a SAHM. Also pregnant." ($P_{759}$)}\\
         33 & Issues with Childcare $(T_{14})$ & daycare, childcare, f*ck, drama, care\\
         \multicolumn{3}{p{15.3cm}}{"Anyone else just… waiting… every day for quarantine call or notice from daycare?" ($P_{1317}$), "daycare dilemma" ($P_{1827}$)}\\
         30 & Feeling Guilty $(T_{15})$ & mom, guilt, feeling, overreacting, tripping\\
         \multicolumn{3}{p{15.3cm}}{"Mom guilt at its finest: my son got scratched by a cat." ($P_{2258}$), "Burn out, mom guilt, wife guilt, etc." ($P_{1401}$)}\\
         \hline
    \end{tabular}
    \caption{Topics discussed in posts by burnt out mothers. Topics are generated using BERTopic and sorted based on frequency.}
    \label{tab:topic_modeling_on_posts_s3}
\end{table}

\subsubsection{What types of comments do posters receive?}
\label{section:study2:content_of_comments}
To better understand comments, we first performed a qualitative coding of 100 randomly sampled comments. When performing this analysis, we were looking to understand the goal of the comment and what it tried to convey. A single comment could have multiple segments aiming to achieve different goals, as a result, a comment could belong to multiple labels. We found that the contents of the comments were made up of one or more of the following six types of content:

\begin{enumerate}
    \item \textbf{Sharing Personal Experiences} $(N = 53)$: A large number of sampled comments shared the commenter's personal experiences. This sharing might be to ensure that the poster does not feel alone. Seeing others going through similar issues could both be encouraging and create a sense of closeness and rapport. When individuals encounter others experiencing similar forms of distress, they often engage in lateral social comparison, which can help normalize their feelings and reduce self-blame~\cite{gerber2018social, buunk2007social}. This sense of shared experience supports communal coping, where the stressor is reframed as a collective challenge rather than an individual failing. Additionally, by sharing personal experiences, commenters not only foster emotional connection but also establish epistemic authority—positioning themselves as credible sources of guidance due to lived expertise~\cite{zagzebski2012epistemic}. This self-positioning is further reinforced by social proof, as others are more likely to accept advice from those who have navigated similar challenges. Moreover, such narratives engage mechanisms of narrative persuasion, making the shared advice more emotionally resonant and persuasive.
    \item \textbf{Emotional Validation or Encouragement} $(N = 47)$: A large portion of the comments offered explicit emotional validation and/or encouragement to the poster. For instance, phrases such as ``oh boy do I feel you.", ``I'm sorry mama", and ``It definitely [...] gets better, but it's completely okay to feel how you're feeling now!" were used in the annotated comments. As evident by this categorization, emotional support is a common response to Reddit posts written by burnt out mothers. 
    \item \textbf{Advice} $(N = 42)$: Our third most comment content type was advice or informational support. These comments often offered exact next steps to do or even what to say to one's partner to solve the issue the poster discussed. 
    \item \textbf{Encouragement to Seek Help} $(N = 10)$: A number of posts encouraged the poster to seek professional help by going to therapy (e.g., ``Get therapy and ask for help"). These comments emphasized that the mother did not need to handle everything on their own, and they should try to find help. 
    \item \textbf{Follow-up Questions (Asking for Additional Information)} $(N = 13)$: A number of comments asked follow-up questions, trying to clarify aspects of the post's content or the specific issue the poster wanted advice or help with. 
    \item \textbf{Rule Reminders \& Resource Sharing} $(N = 6)$: These comments were often automated or template messages reminding the poster and/or commenters of subreddit rules and the tone the subreddit aims to embody. 
\end{enumerate}

It is worth noting that advice-giving was not the most common type of comment. However, the fact that a large number of posts were made with the goal of venting could explain this phenomenon. Additionally, as prior work has pointed out, there is often a ``information-to-application gap" when it comes to parenting advice~\cite{10.1145/3653690}. This gap states that when parents are facing issues, the problem is often not just a lack of knowledge, but rather the ability to develop methods parents are able to implement in their lives~\cite{10.1145/3653690}. Given that our qualitative analysis indicates that the majority of commenters are mothers themselves (inferred by the set of comments that shared their own experiences) they might have an intuitive sense of the lower value of purely providing information. Even when users explicitly seek advice, they often receive responses that do not offer direct advice. Prior work on Reddit has examined this phenomenon and developed a taxonomy to categorize the different types of non-advice responses~\cite{vepsalainen2022responses}.  

To better understand comments at scale, we performed topic modeling on the comments, the results of which are displayed in Table \ref{tab:topic_modeling_on_comments_s3}. Issues between partners, and co-parents $(TC_1)$ was the most commonly commented on topic. This is likely because this was a common issue brought up by posters and a significant issue burnt out mothers deal with. Comments in this group often emphasized the importance of asking for help and communicating with one's partner about difficulties. This issue was closely related to $TC_6$ (\textit{co-parent involvement}). However, in $(TC_1)$ users discussed general dissatisfaction with the partner or how much help they receive in all areas of life. Comments in $(TC_6)$ on the other hand, focused on the day-to-day care of the child (rather than broader relationship issues). 

Discussions about career were also common. While topics $(TC_2)$ and $(TC_{18})$ are both related to one's career, $(TC_2)$ focused more on general career paths and if some careers are more friendly to mothers. While the other $(TC_{18})$ had discussions of how to manage expectations, meetings, and time within the poster's current job. Advice about specific aspects of parenting such as the child's sleep patterns or struggles $(TC_5)$, play time $(TC_8)$, and feeding were also brought up in comments $(TC_{11})$. In the food category for instance, comments talked about what types of foods or exact recipes they used that they thought might be helpful for picky eaters. They also talked about if and when they thought the mother should or should not get worried about a specific eating behavior they were observing. 

\rowcolors{2}{cyan!15}{white}
\begin{table}[]
    \centering
    \begin{tabular}{c p{4.5cm} p{8.5cm}}
        \textbf{Frequency} & \textbf{Topic} & \textbf{Top Keywords} \\\hline    
         6,319 & Between-Partner Struggles $(TC_1)$ & like, get, need, thing, help, time, feel, know, you're, husband\\
         \multicolumn{3}{p{15.3cm}}{"He’s not going to suddenly be the partner you need him to be if he’s not already." ($C_{21,339}$)}\\
         3,243 & Career Path $(TC_2)$ & job, work, time, leave, back, month, year, company, like\\
        \multicolumn{3}{p{15.3cm}}{"So I'm one of those who loves art/photography and did not take a job in those fields. One, photography would require a lot of weekend work. Two, I would lose my passion for it." ($C_{22,063}$)}\\
        2,911 & Self-Care $(TC_3)$ & day, daycare, work, time, hope, husband, kid, week, hour\\
        \multicolumn{3}{p{15.3cm}}{"Therapy. Gym with a daycare room. Journaling. Crying to my mom and husband when needed." ($C_{15,659}$)}\\
        2,683 & Encourage \& Commiserate  $(TC_4)$ & sorry, you're, hope, thank, glad, happy, feel, congratulation, better, post, great\\
        \multicolumn{3}{p{15.3cm}}{"I'm sorry you're going through this right now. It's so hard to read about someone having trouble and not be able to help!" ($C_{1,398}$), "Solidarity momma. I feel like motherhood has ruined my life." ($C_{8,338}$)}\\
         2,634 & Sleep $(TC_5)$ & sleep, baby, month, night, pump, bottle, formula, time, milk\\
        \multicolumn{3}{p{15.3cm}}{"A lot of 2ish year olds attempt to quit napping. It is a RUSE. They still need it. Keep persisting in nap time." ($C_{3,347}$)}\\
        2,121 & Co-parent Involvement $(TC_6)$ & child, kid, parent, want, need, like, life, dad, time\\
        \multicolumn{3}{p{15.3cm}}{"I don't read anything about the father of your child. Is he involved, or are you a single mom?" ($C_{5,911}$)}\\
        1,611 & Rule Reminders, Warnings, \& Automated Messages $(TC_7)$ & reminder, bot, please, rule, post, concern, posing, downvotes, moderator, question\\
        \multicolumn{3}{p{15.3cm}}{"Reminder to commenters: Don't be a prick! Share kindness, support and compassion, not criticism." ($C_{3,937}$)}\\
        1,198 & Play $(TC_8)$ & play, time, kid, day, toy, toddler, thing\\
        \multicolumn{3}{p{15.3cm}}{"Yesterday, we wanted to go camping. That involved me carrying something to his room and then laying down with a blanket on me while he set up his "campsite." " ($C_{14,623}$)}\\
         863 & Work \& Childcare $(TC_{9})$ & sahm, working, home, work, time, job, kid, easier, day\\
        \multicolumn{3}{p{15.3cm}}{"I've been a sahm mom for almost 5 years now [...] I learned early on that I need to get dressed every day, if I lounge in my sleep clothes, then I won't be productive at all." ($C_{9,879}$)}\\
        853 & House Care $(TC_{10})$ & clean, cleaning, laundry, house, meal, husband, help\\
        \multicolumn{3}{p{15.3cm}}{"Writing out the list of things that need to be done and dividing it. Since you’re home more in the early day maybe you can handle laundry and [...]" ($C_{21,201}$)}\\
         535 & Food $(TC_{11})$ & meal, food, cheese, chicken, eat, veggie, cook, sauce, dinner, frozen\\
        \multicolumn{3}{p{15.3cm}}{"Chips and fish fingers. Toast and scrambled eggs, or a tin of baked beans." ($C_{8,246}$)}\\
        482 & Virtual Support $(TC_{12})$ & hug, sending, sorry, you're, bromo, big, virtual\\
        \multicolumn{3}{p{15.3cm}}{"I hope you figure it out. Virtual hugs to you momma." ($C_{5,865}$)}\\
        327 & Schooling $(TC_{13})$ & school, homeschooling, kid, homeschool, day, year, public, teacher, learning, grade\\
        \multicolumn{3}{p{15.3cm}}{"She's either in public school, in private school, or you file a NOI and fully homeschool." ($C_{10,048}$)}\\
        320 & Hired Childcare $(TC_{14})$ & nanny, hire, help, work, time, babysitter, daycare\\
        \multicolumn{3}{p{15.3cm}}{"Can you hire a very part time babysitter? To just cover those half days for the next few weeks?" ($C_{22,134}$)}\\
        282 & Attention \& Gifts $(TC_{15})$ & birthday, Christmas, gift, mother, day, holiday, year, party\\
        \multicolumn{3}{p{15.3cm}}{"All I want is a foot bath. I asked for one. My husband said that that's what our son will get me for mother's day." ($C_{12,224}$)}\\
        247 & Sex Lives $(TC_{16})$ & sex, reward, massage, touched, touch, feel, love, maybe\\
        \multicolumn{3}{p{15.3cm}}{"Sometimes when we get in a funk where we're  just not interested at all, we do it anyway and schedule sex a couple times a week." ($C_{876}$)}\\
        221 &  Need for Support $(TC_{17})$ & village, family, people, help, friend, live, support, parent, mom\\
        \multicolumn{3}{p{15.3cm}}{"Oh love, I had no village either…….it sucks. You are more than allowed to vent your damn head off" ($C_{11,378}$)}\\
        194 &  Work Guidelines $(TC_{18})$ & meeting, work, decline, calendar, hour, time, block, need, day, boss\\
        \multicolumn{3}{p{15.3cm}}{"Decline the meeting and say your workday doesn’t start until x time." ($C_{17,662}$)}\\\hline 
    \end{tabular}
    \caption{Topics discussed in comments in response to posts made by burnt out mothers. Topics are sorted based on frequency. Topics were generated automatically using BERTopic.}
    \label{tab:topic_modeling_on_comments_s3}
\end{table}

Our final step to better understand comments was to categorize them according to the support type they offered (displayed in Table \ref{tab:comment_support_type}). We can see that emotional support is more common in responses than informational support. This finding is inline with our manual coding of comments. The high level of emotional support is also inline with patterns observed in responses to general support seeking messages on Reddit~\cite{adelina2023stories}. Additionally, both types of support are observed to exist in a single comment, confirming our qualitative coding of multiple intents in the same comment. 

To understand the content of comments that were classified as having neither type of support (not emotional or informational), we randomly sampled 20 first-level comments within this category (total = 2,027). A qualitative analysis of these comments revealed that the majority offered short personal stories, stating how the commenter experiences or addresses the issue discussed by the poster (e.g., ``My mom and sister are visiting and we’re ordering a pizza. That’s the amount of planning I want to do."). Other comments within this random subset provided sarcastic or humorous responses (e.g.,``Mercury is in retrograde. Maybe that's their home planet, lol"). Prior work on nursing burnout has shown the use of humor to be a coping mechanism for this cohort~\cite{talbot2000association}. However, the use of humor in the instance of prior work was by the burnt out participant themself. In the comments, however, humor is being used by other's about the poster's circumstances. We are unaware of studies that look into how other people using humor to respond to someone's burnout might affect the person. 


\rowcolors{2}{white}{white}
\begin{table}[]
    \centering
    \begin{tabular}{p{3cm} c c c c c}
        \textbf{Category} & \textbf{Total} &  \textbf{Emotional} & \textbf{Informational}  & \textbf{both} & \textbf{Neither} \\\hline 
        All comments & 50,674 & 34,519 (68.1\%) & 24,559 (48.5\%) & 14,614 (28.8\%) & 6,210 (12.2\%)\\
        First-Level Comments & 27,044 & 19,167 (70.9\%) & 15,887 (58.7\%) & 10,037 (37.1\%) & 2,027 (7.5\%)\\\hline
    \end{tabular}
    \caption{Type of support offered in the comments of maternal burnout submissions on Reddit.}
    \label{tab:comment_support_type}
\end{table}

To better understand the supportive nature of comments, we also looked into the content of comments and its relation to support type. For instance, we found that the majority of comments $(81\%, N = 5118)$ that discuss between-partner struggles $(TC_{1})$ included emotional support. Additionally, 58\% offered informational support, with only 5\% being classified as providing no support. Most other topics had more emotional support being provided (compared to informational or none). However, sleep $(TC_{5})$, play $(TC_{8})$, food $(TC_{8})$, and schooling $(TC_{13})$ had considerably higher (i.e., 10\% difference or more: $(Informational - emotional) > 10\%$ ) informational content than emotional. 

\subsubsection{What is the relationship between seeking and providing support?}
\label{section:reciprocity_finding}Having analyzed the content of posts and comments in the previous sub-sections, we turn to our final question: do mothers who post and seek support and advice, give back and interact with content posted by other members of the community? The norm of reciprocity, is an important concept in all aspects of human life~\cite{gouldner1960norm}. Studies have pointed to the importance of this concept for the stability of social systems, as well as tensions it could introduce to social systems when an individual's return is deemed inappropriate or insufficient~\cite{gouldner1960norm}. Prior work on reciprocity in the workplace has shown that ``vicious cycles'' can be created when employees don't help others because they do not receive help themselves~\cite{deckop2003doing}. These studies motivate our investigation of the existence or lack of reciprocity in the communities burnt out mothers participate in. Understanding how and whether reciprocity occurs in these interactions offers insight into the overall health and sustainability of peer support within these subreddits.

As previously stated, we are analyzing 3,244 posts with the 50,674 comments written in response to these posts. We find that the 3,244 posts, were authored by 2,759 unique users. The 50,674 comments, on the other hand, were authored by 21,816 unique users. To understand if poster $P_i$ engages with other posts made by burnt out mothers, we look for comments written by $P_i$ on posts authored by others users ($\forall_{j \neq i} P_j$). We find that only 612 posters (22\%) interact with other posts within our dataset (i.e., only 22\% interact with posts about the experience of burnout by other mothers). The fact that posters engage in both advice seeking and advice giving is inline with findings from prior wok~\cite{kang2025user}. For instance, female users have been shown to be more likely to be repeat advice-seekers and to engage in advice-giving alongside their future advice-seeking attempts~\cite{kang2025user}. 


\section{Study 3: Understanding Manifestations of Co-Parenting Struggles}
\label{section:study3}
In the content analysis presented in our second study, we found that between-partner issues and co-parenting struggles was something that came up often in Reddit posts (see Table \ref{table:needs_qualitative} and \ref{tab:topic_modeling_on_comments_s3}). 
Co-parenting refers to how parental figures coordinate, support, and share responsibilities in childrearing, and is distinct from the romantic or marital relationship between partners~\cite{xiao2021effects}. The Ecological Model of Co-parenting outlines four key dimensions: division of labor, childrearing agreement, support versus undermining, and joint family management, all of which significantly shape parental mental health and family functioning~\cite{xiao2021effects}. Prior work on parental burnout has also pointed to co-parenting as a potential mitigating factor~\cite{bastiaansen2021gender}. However, this prior work found co-parenting to be more effective in the mitigation of parental burnout for fathers, with a mitigating effect not being found to hold for women~\cite{bastiaansen2021gender}. In this study, we conduct an in depth analysis of co-paring issues as reported by burnt out mothers. While prior work has acknowledged the growing relevance of inter-generational co-parenting—particularly involving grandparents in some countries~\cite{kekkonen2023interpretative, xiao2021effects}—this study limits its scope to traditional co-parenting dyads (e.g., two parents actively sharing caregiving responsibilities), in order to examine how mothers experiencing burnout describe and navigate the dynamics of shared parenting.


\subsection{Method}

To investigate discussions of co-parenting within our set of 3,244 posts, we first needed to extract segments of posts that were relevant to this issue. To extract these segments, we first performed a manual annotation of 100 posts and then developed automated models to perform the extraction of relevant phrases.  

\subsubsection{Co-Parenting Definition} For our investigation of co-parenting issues, we use the definition of co-parenting as ``the means by which parents take the responsibility of the socialization, care and upbringing of the child"~\cite{eira2021co}. We refer readers to the following sources for comprehensive reviews of co-parenting concepts: \cite{mchale2011coparenting, xiao2021effects}. 


\subsubsection{Sampling Strategy} To create a set of posts to label for discussions of co-parenting we first selected a number of words that could be used to discuss this topic. To find these words we performed part of speech (POS) tagging on our dataset of posts written by burnt out mothers (N = 3,244), and manually went through the most repeated nouns to find a list of partner-related nouns. From this list we selected the following partner-related nouns which may be used to refer to the poster's co-parent: \textit{husband}, \textit{boyfriend}, and \textit{dad}\footnote{We acknowledge that our selected keywords limit our labeled dataset (N = 100) of co-parenting issues to those occurring in heterosexual relationship dynamics. However, the same limitation is not placed on the posts analyzed in our findings. The trained co-parenting sentence extraction model is deployed on all 3,244 posts regardless of the existence of any keywords.}. Having selected these keywords, We then randomly sample 100 posts from the set of posts that included at least one of these words. 

\subsubsection{Manual Annotation} To create our dataset of co-parenting sentences, we conducted manual labeling of the 100 posts (sampled using the method described above). Two annotators (one graduate computer science student and one undergraduate computer science student with prior experience in data annotation) performed this task. Prior to performing the labeling, the annotators discussed the definition with each other in a meeting. Each annotator was then tasked with extracting all sentences from each of the 100 posts that discussed co-parenting issues based on the definition. At the end of the initial annotation phase, the annotators reached a moderate level of inter-annotator agreement, as measured by the Jaccard Index—a widely used metric that quantifies the degree of overlap between two sets by dividing the size of their intersection by the size of their union~\cite{jaccard1901etude}. Given the limited alignment at this stage, the annotators engaged in an iterative, collaborative review. During this phase, they revisited each sentence, discussed their interpretations, and resolved discrepancies by examining overlooked aspects and converging on a shared understanding. Following this process, inter-annotator agreement substantially improved, reaching a high level of consistency (Jaccard Index = 0.92). The final disagreements between the annotators were then resolved by a third expert annotator (a computational social science expert with 10+ years of experience). 55\% of posts in the labeled dataset had at least one sentence related to co-parenting, with the average number of sentences per post being 0.95. This average number of sentences per post for the set of posts that discuss co-parenting (N = 55) was 1.72. 

\subsubsection{Automatic Extraction} To efficiently extract sentences across all 3,244 posts, we developed automated systems and systematically experimented with different prompting strategies and LLMs. Specifically, we explored two prompting strategies: an extraction-focused prompt, which directly instructed the model to extract relevant content, and a question-answer format, which framed the task as a question-answering problem. We applied the Automatic Prompt Optimization method~\cite{jm3} to iteratively search for prompts with improved performance. For the extraction-focused prompt, we began by including the exact definition of co-parenting in our instructions. We used the F1 score as a scoring metric to assess whether the prompt's performance improved and employed GPT-4o as an agent for prompt expansion and variation generation. This process was conducted for two iterations to obtain the final optimized prompt. For the question-answer format prompt, the initial prompt consisted of a system prompt containing only the base definition without any additional instructions. During subsequent iterations, we incorporated rules of inclusion and exclusion to refine the final optimized prompt. The final forms of these prompts are presented in Table \ref{tab:coparenting_prompts}. These strategies were tested under zero-shot and five few-shot settings (zero, two, four, six, and eight exemplars), selected based on prior work showing that adding more in-context examples beyond a small number offers limited performance gains~\cite{min2022rethinking}. For the few-shot examples, we employed the bert-base-nli-mean-tokens model~\cite{reimers-2019-sentence-bert} to generate sentence embeddings and used cosine similarity to select the closest examples to each Reddit post, following an in-context retrieval strategy~\cite{liu2021makes}.

\begin{table}[h]
\centering
\renewcommand{\arraystretch}{1.3}
\begin{tabular}{|p{0.47\textwidth}|p{0.47\textwidth}|}
\hline
\textbf{Prompt 1: Extraction-focused Format} & \textbf{Prompt 2: Question-answer Format} \\
\hline

\textbf{System Prompt:} \newline
You are tasked with identifying quotes about co-parenting from social media posts about burnout. You will be given a Reddit post and must extract the exact sentences that talk about the coparenting of the child, with respect to the poster's coparent.

Co-parenting involves parents who together take on the socialization, care, and upbringing of children for whom they share equal responsibility. \textbf{Exclude} any references to the poster's own parents or other family members. Extract only sentences that discuss the co-parent's \textbf{involvement, presence, or absence} in the child's life. \textbf{Do not include} sentences that focus solely on the relationship between the poster and their co-parent unless it directly affects the child. \textbf{Exclude} any mention of pregnancy, unborn children, or non-biological parents unless they are explicitly co-parenting. Do \textbf{not} include quotes about the poster's own parenting.

Output format: Return a list of full sentences. If there are none, return \texttt{NONE}.

Example Output: [``Sentence\_1'', ``Sentence\_2'']

\textbf{User Prompt:} \newline
\texttt{Reddit Post: \{Reddit Post\}} &

\textbf{System Prompt:} \newline
You will be given social media posts about burnout. You need to answer the question related to it.

Co-parenting refers to parents who together take on the socialization, care, and upbringing of children for whom they share equal responsibility. Do not include references to the poster's own parents or other family members.

\textbf{User Prompt:} \newline
Question: Which statements in the post does the author talk about their coparent with respect to their child? Do not mention sentences that only describe the relationship. If there are no such mentions, return \texttt{None}. A coparent cannot be the poster's parent or the child's grandparent. Do not include unrelated sentences. Return full sentences only.

Example Output: [``Sentence\_1'', ``Sentence\_2'']

\texttt{Reddit Post: \{Reddit Post\}} \\
\hline
\end{tabular}
\caption{Comparison of two prompt formulations for extracting co-parenting references from burnout-related Reddit posts.}
\label{tab:coparenting_prompts}
\end{table}

Table~\ref{tab:coparenting_f1} presents the F1 scores across different settings. We initially tested GPT-4o~\cite{openai2024gpt4technicalreport}, a widely adopted model known for achieving state-of-the-art performance across various benchmarks. Despite the extraction-focused prompt providing more detailed instructions, the question-answer format consistently achieved better performance, with a peak F1 score of 0.72. For the extraction-focused prompt, the highest F1 score was obtained using two exemplars, but performance decreased with additional examples. In contrast, for the question-answer format, adding exemplars significantly reduced performance from 0.72 to around 0.52. To further explore model performance, we tested DeepSeek-R1~\cite{deepseekai2025}, known for its strong reasoning capabilities, and Llama-3.1-8B~\cite{grattafiori2024llama3herdmodels}, a robust open-source model commonly used in research, using the best-performing settings for GPT-4o (question-answer format with zero-shot setting). As shown in Table~\ref{tab:coparenting_f1}, both DeepSeek-R1 and Llama-3.1-8B underperformed compared to GPT-4o. Based on these results, we selected the zero-shot GPT-4o model with the question-answer format prompt for sentence extraction from the entire dataset.

\begin{table}[h]
\centering
\renewcommand{\arraystretch}{1.3}
\begin{tabular}{|l|c|c|c|c|c|}
\hline
\textbf{Setting} & \textbf{0-shot} & \textbf{2-shot} & \textbf{4-shot} & \textbf{6-shot} & \textbf{8-shot} \\
\hline
GPT-4o-Extraction-focused & 0.49 & 0.53 & 0.44 & 0.44 & 0.46\\
\hline
GPT-4o-Question-answer format & \textbf{0.72} & 0.53 & 0.51 & 0.52 & 0.52\\
\hline
DeepSeek-R1-Question-answer format & 0.51 & - & - & - & - \\
\hline
LLaMA-3.1-8B-Question-answer format & 0.46 & - & - & - & - \\
\hline
\end{tabular}
\caption{F1 scores across different models and prompting strategies with varying few-shot KNN settings for the detection of co-parenting sentences. The highlighted cell indicates the highest performing model.}
\label{tab:coparenting_f1}
\end{table}

\subsubsection{Analysis} Using the best GPT model (see Table \ref{tab:coparenting_f1}), we extracted all co-parenting sentences from the 3,244 posts predicted to be written by burnt out mothers. A total of 1,177 (36\%) had at least one extracted sentence. From these 1.1K posts, 3,366 separate co-parenting sentences were extracted. The distribution of sentences across the 1.1K posts were as follows: $min = 1$, $mean = 2.86$, $median = 2$, $max = 87$. To decide whether sentences should be analyzed individually or grouped at the post-level, we sampled 100 sentences (from the set of 3,366 sentences) and had one annotator (a graduate computer science student with prior experience in annotation) go through them manually, indicating if the sentence stands on its own or if additional context is required. The annotator found that 82 of the sample of 100 stood on their own, while others required additional context. Given that a large portion stood on their own, we used sentences as our unit of analysis, using them individually for the remainder of our analysis. 

Our first analysis to better understand discussions of co-parenting was to perform topic modeling on the extracted sentences. Similar to previous studies we used both BERTopic~\cite{grootendorst2022bertopic} and LLooM~\cite{lam2024conceptInduction} for this study as well. However, a manual analysis of the results of both models revealed that results of LLooM were more informative in this instance. Results of BERTopic in this case were too high-level and not informative enough. The results of LLooM (as presented in Table \ref{tab:co_parenting_topic_modeling}), on the other hand, presented meaningful sub-topics and dynamics of co-parenting. It is important to note that a single sentence could be classified in multiple categories by the LLooM model. LLooM originally outputted 14 topics. However, we decided to discard four of these topics because they were too generic, and were identified in the majority of sentences. The removed topics were as follows: \textit{Parental Burnout}, \textit{Parenting Challenges}, \textit{Parenting Duties}
, and \textit{Childcare Management}. We present the other 10 topics in this study's findings.  

To better understand co-parenting mentions, we also performed a qualitative thematic analysis of these sentences. To do so we sampled 100 sentences and one member of our research team (a graduate computer science student with prior experience in qualitative research) went through these sentences and performed inductive coding on them. The goal of this step was to annotate how burnt out mothers talked about their co-parent and what issues they brought up. A total of 9 codes were discovered. We present the results of this analysis in the findings of this study.

\subsection{Findings}

\subsubsection{What aspects of co-parenting are burnt out mothers discussing?} Table \ref{tab:co_parenting_topic_modeling} displays the prominent areas of concern and discussion extracted using LLooM. As observed in the table, issues relating to bonding between child and parent/co-parent $(CPT_{1})$ is the most common topic. The prevalence of this topic indicates that mothers care a great deal about the parent-child relationship. Looking through examples fitting within this category, we find that discussions of this topic include both positive and negative examples of bonding. In the example shown in the table we are seeing a positive example of bonding between the co-parent and child, praising the co-parent for how involved they are in the upbringing of the child. However, negative instances of the same phenomena were also observed, in which the mother was  dissatisfied with the low effort put in by the co-parent. The prominence of bonding-related concerns in mothers' posts reflects the salience of the parent-child relationship in maternal identity and emotional life~\cite{johnson2013maternal}. Grounded in attachment theory~\cite{howe2012attachment} and parenting psychology, this salience manifests in both affirming narratives (e.g., praising involved co-parents) and expressions of distress when bonding is perceived as lacking or insufficient. This topic could be impacted by or be associated with some of the other topics displayed in the table such as: \textit{Husband's Work Impact $(CPT_{2})$}, \textit{Husband's Domestic Role $(CPT_{5})$}, and \textit{Husband's Leisure Time $(CPT_{10})$}. For instance, a number of mothers discussed how their co-parents' work schedule made it hard for them to coordinate and/or split tasks $(CPT_{2})$. Within the discussion of work schedules, instances of negotiating to change said work schedule was also brought up. 

Arrangements between parents who were no longer together were also common $(CPT_{3})$. These parents, who were divorced or separated, needed to manage visitation schedules, with frustrations coming up when this scheduling was not done properly. Financial issues $(CPT_{8})$ were also observed. Within these sentences, mothers discussed if their family was facing financial struggles, if they worked themselves, and if the co-parent was able to financially provide for the family or not. When their co-parent was the financial supporter of the family, mothers expressed feeling like they were not able to spend money on things they wanted or cared about. These findings align with burnout research showing that a lack of control, unmet expectations, and insufficient support—whether emotional, logistical, or financial—can exacerbate feelings of entrapment and chronic stress in parenting roles~\cite{mikolajczak2019parental, lindstrom2011parental}.

\rowcolors{2}{orange!15}{white}
\begin{table}[]
    \centering
    \begin{tabular}{c p{5cm} p{8cm}}
        \textbf{Frequency} & \textbf{Topic} & \textbf{Inclusion Criteria} \\\hline  
         2,928 & Parent-Child Relationship $(CPT_{1})$ & Does the text detail interactions or activities that affect the relationship and bonding between a parent and child?\\
         \multicolumn{3}{p{15.3cm}}{"He's so affectionate and hands on with both our children." ($S_{3275}$)}\\
         2,766 & Husband's Work Impact $(CPT_{2})$ & Does the text describe how the husband's work schedule or job responsibilities affect his family or parenting duties?\\
         \multicolumn{3}{p{15.3cm}}{"Hes already agreed to find new work locally where he can be home every day" ($S_{36}$)}\\
         2,559 & Custody and Visitation $(CPT_{3})$ & Does the text discuss issues related to custody, visitation rights, or co-parenting arrangements?\\
         \multicolumn{3}{p{15.3cm}}{"Last night her dad tells me that BIOMOM has unilaterally decided to take her kids trick or treating" ($S_{62}$)}\\
         2,345 & Emotional and Behavioral Issues $(CPT_{4})$ & Does this text describe emotional or behavioral issues faced by children or parents?\\
         \multicolumn{3}{p{15.3cm}}{"My husband is more patient than me and doesn't seem to be experiencing this so of course I feel like a terrible mother." ($S_{1711}$)}\\
         2,325 & Husband's Domestic Role $(CPT_{5})$ & Does the text discuss the husband taking on domestic or childcare responsibilities?\\
         \multicolumn{3}{p{15.3cm}}{"my husband is fine but will any man ever be mom? He expects a lot of invisible labor of me" ($S_{3092}$)}\\
         2,240 & Parental Support Dynamics $(CPT_{6})$ & Does the text describe how one parent supports or fails to support the other in parenting?\\
         \multicolumn{3}{p{15.3cm}}{"Something my husband *forgets* to feed him." ($S_{1716}$)}\\
         2,063 & Child Safety Concerns $(CPT_{7})$ & Does this text involve concerns or incidents related to the safety or well-being of a child?\\
         \multicolumn{3}{p{15.3cm}}{"It feels like a constant tango of tension, and we're frequently bickering in front of our daughter and it breaks my heart to expose her to that." ($S_{699}$)}\\
         752 & Husband's Financial Role $(CPT_{8})$ & Does the text discuss the husband's role in providing financial support or managing finances for the family?\\
         \multicolumn{3}{p{15.3cm}}{"Husband makes over 100k yet he says we still can't afford preschool." ($S_{858}$)}\\
         624 & Husband's Mental Health $(CPT_{9})$ & Does the text mention the husband struggling with mental health issues and how it affects his family or parenting?\\
         \multicolumn{3}{p{15.3cm}}{"After a long talk with my husband it appears he's burnt out on watching our 11 month old." ($S_{1103}$)}\\
         363 & Husband's Leisure Time $(CPT_{10})$ & Does the text describe the husband spending time on leisure activities or personal interests?\\
         \multicolumn{3}{p{15.3cm}}{"He'll also sit around watching videos on his phone at full volume while I'm trying to feed/clean up after our baby, totally zoned out." ($S_{2341}$)}\\\hline 
    \end{tabular}
    \caption{Topics discussed in the extracted co-parenting sentences detected using LLooM~\cite{lam2024conceptInduction}. Topics are ordered by frequency.}
    \label{tab:co_parenting_topic_modeling}
\end{table}

\subsubsection{Qualitative Analysis of Co-parenting sentences} Through inductive coding, we extracted nine patterns when it came to how burnt out mothers talked about their co-parents. We will go over these different patterns in this section.

\noindent\textbf{1: Appreciating One's Partner}: While many posts expressed frustration with unequal parenting dynamics, a subset of mothers described their partners in highly positive terms, highlighting appreciation for their involvement and support. These posters emphasized that their co-parents were emotionally present, actively engaged in childcare, and contributed meaningfully despite personal or situational challenges. One mother wrote, ``My partner helps as much as humanly possible, and he has severe chronic health problems so I'm shocked he hasn't dropped dead of a heart attack from stress" ($Sample_{94}$). Others described their partners as deeply invested in parenting: ``He is a truly great parent and he does so much with her" ($Sample_{99}$), and ``My husband is great he does everything he can" ($Sample_{89}$). Emotional availability and shared responsibility were also noted, with posters expressing ``He is emotionally available, takes interest in our kid, works hard to provide for our family" ($Sample_{73}$), and ``And just to add [my partner] is a wonderful and incredible dad" ($Sample_{85}$). These accounts, while less frequent, suggest that active and empathetic co-parenting is very important to and appreciated by burnt out mothers. Research on parental burnout shows that perceived emotional and instrumental support is a key protective factor that mitigates emotional exhaustion and fosters resilience~\cite{findling2024parental, lebert2022does}.






\noindent \textbf{2: Child-Partner Bonding (or Lack thereof)}: Some posts focused on the relationship between the child and the co-parent, highlighting both moments of connection and instances of strain. In positive accounts, mothers described growth in the emotional bond between the child and their partner, often attributing this to increased involvement in caregiving. As one poster noted, ``This also seems to have helped the kids bond more with dad, which he loves" ($Sample_{70}$). Another shared a meaningful milestone: ``[co-parent] won [child's] heart and is now DADA" ($Sample_{78}$). However, not all posts reflected positive relational dynamics. Some mothers expressed concern about their partner's emotional approach to parenting, particularly when it risked alienating the child. One poster wrote, ``In a way he thinks he's showing her that if she's horrid he'll be cold, but I've had a stern word with him that you can't do that and it'll drive a wedge between them" ($Sample_{14}$). These examples underscore the significance of co-parent–child bonding in shaping the overall parenting dynamic, and the role mothers often play in mediating or facilitating that relationship. This topic was also reflected in our automated topic detection results $(CPT_{1})$.




\noindent\textbf{3: Unequal Expectations \& Treatment}: Mothers reflected on the broader societal and internalized expectations that shaped their co-parenting experiences, often highlighting how caregiving responsibilities defaulted to them regardless of relationship structure or employment status. Even in households that outwardly aspired to egalitarianism, the burden of managing the home and children frequently remained with the mother. As one user shared, ``Even when my husband is around, I am usually the one who wakes up to tend to my son" ($Sample_{90}$). These unequal expectations extended beyond the home and into how others perceived parental roles. For example, one mother noted, ``Interestingly when my husband was let go in June [...] no one asked if he was going to be spending all this extra time with [our child]" ($Sample_{91}$). In some cases, mothers described a stark asymmetry in caregiving standards for themselves versus their partners: ``Keep in mind he does not take on ANY parental roles or bonding when it comes to my daughter but has every expectation that I will for his" ($Sample_{63}$), and ``My SO wants me to be there with his kids all the time" ($Sample_{84}$). Tensions around unequal treatment were also present in professional contexts, where mothers felt their careers were devalued in comparison to their partner's: ``My job is just as important to me as yours is to you!" ($Sample_{59}$). These accounts illustrate how persistent gendered assumptions around caregiving continue to influence co-parenting dynamics and contribute to maternal burnout. Perceived inequity in parental roles is a central driver of parental burnout, especially when caregiving demands are high and unreciprocated~\cite{roskam2020gender}.






\noindent\textbf{4: Prioritizing Self \& Neglecting Partner's Needs}: Some mothers described feeling like their needs were overlooked by their partners, particularly in moments when support was most needed. These posts expressed frustration at partners who routinely prioritized personal leisure or rest over shared caregiving responsibilities. One mother recounted, ``He often goes out to golf or hang with his friends without even considering me/the children and what our needs are" ($Sample_{65}$), highlighting the emotional disconnect and lack of coordination in co-parenting. Another poster similarly expressed dismay at the imbalance in daily responsibilities: ``You know what he's currently doing while I'm sitting on our couch holding our baby? Sleeping" ($Sample_{15}$). These accounts reflect a pattern in which mothers felt their own needs were consistently subordinated—not only by the demands of parenting, but by partners who failed to reciprocate the labor of care. Prior research shows that a lack of leisure time and the inability to recover from daily stressors significantly contributes to parental burnout~\cite{lindstrom2011parental}.



\noindent\textbf{5: Weaponized Incompetence}: A recurring theme in some posts was the perception that co-parents exaggerated their inability to manage parenting tasks as a means of avoiding responsibility—a phenomenon often described as ``weaponized incompetence." Mothers expressed frustration at partners who deflected caregiving duties by positioning themselves as inherently less capable. One poster shared, ``he claims, every time, that he can't quiet our son because he just wants me" ($Sample_{19}$), suggesting that the partner's self-dismissal was used to shift the burden of taking on the task. Another mother described returning home after a brief absence to find the household in disarray: ``He watches the kids for like 3 hours, and I come home and the house is a disaster!" ($Sample_{49}$). These accounts reflect a dynamic in which the unequal distribution of care work is sustained not through outright refusal, but through a pattern of strategic under-performance that ultimately reinforces maternal overload.



\noindent\textbf{6: Role Imbalance Due to Work Commitments \& Schedules}: Several mothers described how their partner's work schedules or professional demands limited their ability to participate in caregiving, often leading to a persistent imbalance in parenting responsibilities. In these accounts, the burden of daily routines, especially those tied to fixed times like mornings or evenings, frequently fell on the mother. One user explained, ``My husband's commute/hours require him to be out of the house very early in the morning, so our son's morning routine is entirely on me" ($Sample_{8}$). Others highlighted ongoing tensions about how work was used to justify a lack of engagement at home: ``We have this recurring fight about how he can't do a lot at home due to his job" ($Sample_{44}$). While these posts acknowledged external constraints, they also conveyed a sense of frustration, particularly when professional obligations were used to excuse disengagement from shared caregiving duties.



\noindent\textbf{7: Undermined and Judged by the Co-Parent}: Mothers described experiences in which their co-parents undermined their confidence by criticizing or diminishing their parenting, particularly during moments of stress or emotional exhaustion. These posts conveyed a sense of being judged rather than supported, contributing to feelings of guilt and self-doubt. One mother wrote, ``Any time I lose my patience with our offspring he acts as if I am a monster" ($Sample_{33}$), reflecting how minor lapses in composure were met with outsized condemnation. Another expressed frustration at being the target of ongoing criticism: ``he starts subtly attacking me and my mediocrity" ($Sample_{29}$). These dynamics intensified the emotional strain of caregiving, potentially also reinforcing internalized narratives of inadequacy.



\noindent\textbf{8: Being Cast as the ``Not-Fun'' Parent}: A number of posts illustrated the tension that arises when one parent—typically the father—assumes the role of the ``fun parent," leaving the mother to enforce rules and maintain structure. This dichotomy often created resentment, as mothers were positioned as the disciplinarian while the co-parent avoided conflict and responsibility. One user described a pattern of permissiveness in her co-parent's household: ``[co-parent] let them choose what to eat [...], no routine, no boundaries, zero guidance for life, zero savings, sitting up until 3am when they have school the next day" ($Sample_{3}$). Another mother noted the contrasting perception her children had of their parents: ``While they get the well rested, cool, calm and collected dad who's the fun parent" ($Sample_{97}$). These accounts underscore the emotional and logistical toll of being the default authority figure, a role burnt out mothers often carry alone. Research on parental burnout highlights that when positive parenting experiences are monopolized by one partner, the other may experience diminished parental fulfillment---a core component of burnout~\cite{roskam2017exhausted}. 



\noindent\textbf{9: Stress from Poor Co-Parenting Coordination}: Mothers shared stories of important decisions being made without prior discussion, leading to confusion or conflict. For instance, one mother recalled, ``Last night ex emails that he's set up eye exams for the kids... for TODAY" ($Sample_{35}$), highlighting the lack of coordination. In other cases, unclear boundaries around decision-making roles created tension among involved caregivers. One burnt out mother described a situation with a step-parent dynamic: ``They are her children, but that wasn't her time with them to make the parenting decisions and since I'm not the actual parent I didn't say anything" ($Sample_{34}$). These examples illustrate how insufficient coordination can disrupt routines, strain relationships, and complicate the caregiving environment. Burnout literature suggests that such disorganization contributes to a sense of chaos and lack of control, both of which are stressors that heighten burnout risk~\cite{lindstrom2011parental, roskam2017exhausted, glass1996perceived}.



Our analysis of co-parenting dynamics revealed that burnout is shaped by how caregiving responsibilities are negotiated—or left unaddressed—between partners. These findings align with parental burnout theory, which identifies lack of support and perceived unfairness in role distribution as key contributors to exhaustion and emotional distancing~\cite{mikolajczak2018theoretical}. They also extend feminist theories of emotional labor and intensive mothering~\cite{mikolajczak2018theoretical}, illustrating how structural and relational inequities are experienced at the granular level of daily parenting. In particular, mothers' narratives often reflected a dual burden: they were not only performing the majority of childcare tasks, but also managing the emotional consequences of their partner's under-involvement.

\section{Discussion}
\label{section:discussion}
We discuss the implications of our work along three axes. First, we situate our findings within prior investigations of online self-disclosure (Section~\ref{discussion:online_self_disclosure}). Next, we offer a critical reflection on the automated detection of motherhood (Section~\ref{discussion:more_models}). Finally, we conclude with design recommendations for online peer discussion platforms (Section~\ref{section:discussion:design_implications}).

\subsection{Gendered Online Self-Disclosure}
\label{discussion:online_self_disclosure}

Self-disclosure is defined as ``revealing personal or private information about self that is generally unknown and not available from other sources''~\cite{dindia2011online}. When discussing self-disclosure in face-to-face interactions, prior work has often focused on ``verbal'' expressions of the self, rather than nonverbal cues~\cite{dindia2011online}. Online self-disclosure, however, could include numerous instances of nonverbal cues such as the sharing of links or images. Self-presentation is a closely related concept, defined as ``selectively presenting aspects of oneself to control how one is perceived by others and is concerned with impression management''~\cite{dindia2011online, goffman2023presentation}. 

When considering the discourse of mothers online, it is worth noting how the users might be presenting and signaling information about themselves~\cite{donath2007signals}. Strategic self-presentation on social media has previously been studied~\cite{10.1145/2818048.2819927, 10.1145/2702123.2702205, 10.1145/1316624.1316682, 10.1145/3473043, 10.1145/3617654}. For instance, prior work has investigated users' self-presentation and curation practices, as well as others reactions to it, on platforms such as Instagram --- revealing viewer's negative reactions to the excessive use of selfies~\cite{hong2020you}, poster's portrayal of an idealized self~\cite{harris2019instagram}, and user's sharing of content that makes them appear interesting and well liked~\cite{yau2019s}. Unlike Instagram where users often have a dedicated account they use over a long period of time, thus making a history of themselves available to the viewers, Reddit users often use throwaway or multiple account. The use of throwaway or multiple accounts is common when it comes to the discussion of stigmatized topics (e.g., mental health), however, prior work has showcased the use of these practices for impression management~\cite{wohn2021many}. 

Even though the audience of these posts are presumably other parents (given the subreddits they are posted to), parents have been shown to be judgmental towards each other~\cite{10.1145/2818048.2819927}. Given parents can be subjected to scrutiny for their choices, even when using throwaway accounts~\cite{ammari2019self}, they might narrate and disclosure their struggles strategically. Our results, for instance, showed how parents iterate the labor they engage in, state a fear of sharing struggles offline, include a pre-emptive defense of themselves (e.g., I'm not a bad mother) or a strong negative framing of their own competence, as well as at times framing their posting as ``just venting''. Future work should investigate the reasons behind some of these narratives and framings through interview studies. 

Posts framed as venting often did not seem to be written with the primary intention of soliciting advice but instead served as raw disclosures of burnout, exhaustion, anger, and grief. In many cases, mothers explicitly stated that they ``needed to get it out somewhere,'' underscoring the role of online platforms as outlets for emotional expression. This type of venting serves several important functions. Drawing on psychological theories of emotion regulation, expressive writing has been shown to alleviate stress by enabling individuals to process difficult emotions and organize their thoughts~\cite{smyth2003focused}. In venting posts, mothers often voiced thoughts they felt unable to express offline, thus potentially positioning these disclosures as small acts of resistance against the idealized norms of ``good motherhood." Yet, venting posts also revealed tensions between expression and reception. While some readers responded with empathy or validation, others misinterpreted these posts as requests for advice or correction. This could indicate that the imagined and actual audiences of content diverge~\cite{davis2014context}, leading to misaligned expectations about appropriate responses. This ambiguity may signal a need for design interventions that honor venting as a valid form of engagement. 



\subsection{Maternal Health Language Technologies}
\label{discussion:more_models}

In our first study (Section \ref{section:study1}), we developed models to detect self-disclosures of motherhood in online discourse. Accurately identifying mothers within large-scale social media datasets is crucial for enabling deeper investigations into the unique challenges they face. Despite the increasing attention to demographic inference in NLP, there is an absence of tools for detecting maternal identity, such as classifiers that can reliably identify whether a social media post was written by a mother. This gap compelled us to develop our own self-disclosure classifier. This lack of foundational NLP infrastructure stands in stark contrast to the critical importance of maternal well-being as a global health priority. Supporting maternal health is central to several United Nations Sustainable Development Goals (SDG 3: Good Health and Well-being, SDG 5: Gender Equality, and SDG 10: Reduced Inequalities)~\footnote{\url{https://sdgs.un.org/goals}}, yet NLP research has only just begun to consider the specific linguistic needs and expressions of mothers in digital contexts. Our work contributes to this emerging area and demonstrates the value of modeling maternal self-expression as a step toward scalable, real-time insights into caregiver burnout and well-being.

\noindent\textbf{Benefits.} Automating the detection of maternal disclosure on online platforms enables the study of caregiving, identity work, and online support at scale. Maternal disclosures are often deeply personal and tied to experiences of stigma and vulnerability \cite{lebert2018maternal, hubert2018parental}. Online platforms like Reddit are frequently used by mothers who lack adequate offline resources or who face stigma in seeking support through formal channels~\cite{francisco2024hey, pilkington2021mothers, ammari2019self}. Detecting these disclosures computationally allows researchers to move beyond small-scale qualitative coding towards understanding broader patterns of how mothers seek support, provide care for others, and engage with digital communities. Tools that allow automated discussion could be leveraged to create features that surface empathetic responses, flag urgent disclosures, or mitigate harmful interactions~\cite{he2022effects, garg2021detecting}. These types of computational approaches could also addresses several methodological challenges in maternal health research such as inclusion of previously excluded populations (e.g., due to time constraints or geographic barriers), delivery of real-time support, and study of authentic expressions.


\noindent\textbf{Drawbacks.} These benefits, however, come with trade-offs that may undermine trust and cause harm. During automatic detection of motherhood, misclassification poses particular risks: false positives (e.g., mistakenly detecting maternal disclosure) may cause confusion, deteriorate user experience, and reduce trust. False negatives, on the other hand, may overlook individuals who need support~\cite{nguyen2024using}. In addition, it can reduce privacy and expose personal information; it may inadvertently reveal sensitive data to unintended audiences such as governments, advertisers, or employers. For instance, users may not want to be labeled~\cite{pinto2009reducing} or grouped, nor to receive targeted information or resources they did not request, which could lead to intrusiveness or discrimination~\cite{lee2019algorithmic}. Finally, contextual appropriateness is another concern, as the system must carefully consider when, where, and how support is delivered after motherhood detection to avoid potential intrusiveness and harm \cite{hou2025mitigating}. Balancing these opportunities and risks underscores the need for ethically informed and sustainable deployment of automated models to detect maternal burnout.

\subsection{Design Implications}
\label{section:discussion:design_implications}


\subsubsection{\textcolor{red}{Family-Centered Dashboards}} 
As discussed in Section~\ref{section:study3}, posters report imbalances in parenting labour and contributions to parenting tasks. These insights point to the need for interventions that support not just individual coping, but also relational equity through tools that facilitate task-sharing, communication, and reflective co-parenting. This suggestion builds on prior work on family coordination and domestic labor management. This work has explored how families use digital tools to coordinate schedules~\cite{10.1145/1502800.1502806}, manage household tasks~\cite{lee2025development}, and negotiate caregiving responsibilities~\cite{sala2021mobile}. Our findings point to the need for dashboards that extend beyond task management to account for the invisible labor underlying maternal burnout (e.g., emotional and mental load of planning and decision making).


Designing such systems, However, requires careful attention to power dynamics and potential for surveillance. Labor tracking instruments could lead to judgment of partners rather than support. Thus, family-centered dashboards need to be designed to promote partnership rather than policing, potentially through features that emphasize mutual support, celebrate contributions, and frame equity as a shared goal rather than a zero-sum competition. Additionally, these tools should acknowledge the structural constraints that contribute to labor imbalances, such as workplace policies, extended family expectations, or financial pressures.

\subsubsection{Light-Weight Nudges for Community Sustainability} 

In Section \ref{section:study2}, we examined whether users who share their own experiences of maternal burnout also engage with others by commenting on posts made by fellow mothers discussing similar experiences. We found that only 22\% of posters engage with others' posts about maternal burnout, suggesting a relatively low level of reciprocity within the community. From a psychological perspective, this has important implications. According to Social Exchange Theory~\cite{cook2013social, cropanzano2005social} and Norms of Reciprocity~\cite{gouldner1960norm}, mutual interaction—such as commenting on others' posts—fosters a sense of belonging and reinforces ongoing participation. In the context of burnout, however, mothers may lack the emotional or cognitive bandwidth to reciprocate support due to their own depleted resources~\cite{mikolajczak2020parentalIs}. The lack of reciprocal engagement may hinder the development of communal coping~\cite{lyons1998coping} or relatedness which are core components of resilience and recovery in peer support systems. 

Building on existing personalized recommendation systems~\cite{qian2013personalized}, design-based solutions could be developed to encourage users to offer support and advice to others. While it is essential not to pressure vulnerable users into emotional labor they may not be ready for, platforms could explore gentle, opt-in mechanisms to foster community reciprocity. For instance, after writing their own post, users could be shown a small number of similar posts to engage with. Additionally, platforms could offer badges or feedback for empathetic commenting, reinforcing prosocial behavior~\cite{yanovsky2021one, burtch2022peer}. Algorithmic curation could further prioritize content that aligns with a user's lived experience, creating opportunities for low-effort, high-impact support exchange.

Another potential design intervention is the recommendation of a comment draft generated based on the mother's own post. As noted in our findings (see Section \ref{section:study2:content_of_comments}), users frequently share personal stories and struggles with burnout in the comment, possibly as a way of establishing emotional connection with others. An AI-generated draft, drawing from the commenter's own experiences taken from posts they have made, could suggest a relevant and empathetic comment aligned with the content of another mother's post. Providing such a draft may reduce the cognitive burden associated with initiating a response, offering users a concrete starting point and encouraging engagement. This suggestion follows existing research into AI-generated drafts of responses in contexts such as emails~\cite{thiergart2021understanding}, messages~\cite{garcia2024artificial}, and contracts~\cite{wolff2024preparing}. However, it is important to note that existing drafting systems have not yet demonstrated significant success. For instance, prior work found that the utilization rate of AI-generated responses to health messages among clinicians was only 20\%, with no observable reduction in reply time~\cite{garcia2024artificial}. The authors hypothesized that editing AI-generated drafts may be less cognitively demanding even if it still requires as much time as composing responses from scratch~\cite{garcia2024artificial}. Nonetheless, insights from this literature could inform the design of drafting systems that promote reciprocity on social media platforms; for broader AI uses in peer support groups, see~\cite{10.1145/3711089}.



\subsubsection{Allowing for Self-Reflection}
Prior HCI research on reflection interfaces offers guidance for how platforms can better support emotional expressions such as venting~\cite{10.1145/2839462.2839466, 10.1145/3517233}. Reflective design has identified four key principles for supporting meaningful reflection through technology: enabling users to revisit past experiences, providing structured prompts for deeper thinking, facilitating pattern recognition across time, and supporting goal-setting and progress tracking~\cite{10.1145/3517233}. These principles align well with the needs expressed by mothers in our dataset, who often sought to understand their experiences, validate their feelings, and identify paths forward. Following these guidelines, future work could incorporate mood-tracking components that accompany free-text entries to promote emotional awareness~\cite{hamre2021mood}.

Additionally, research on therapeutic writing and expressive journaling has shown that guided prompts can be more effective than free-form writing for emotional processing~\cite{baikie2005emotional}, particularly when they encourage perspective-taking or meaning-making rather than simple venting. This includes structured prompts that gently scaffold reflection—for example, ``What are you feeling right now?" or ``What do you want to remember about this moment?"—as well as the option to set goals or revisit past entries to identify patterns or growth~\cite{10.1145/3517233}. These features could be integrated into maternal support platforms to facilitate private or semi-private expression and allow users to control the visibility and expected response to their posts.



\section{Limitations}
\label{section:discussion:limitations}

While this study offers valuable insights into maternal burnout as expressed in online communities, several limitations must be acknowledged. First, our dataset is limited to Reddit posts, which may not be representative of all mothers experiencing burnout. Reddit users are typically younger and more technologically adept, and the platform's norms may shape the way burnout is articulated. Second, although we employed a robust annotation process and fine-tuned machine learning models to gain a deeper understanding of the challenges faced by mothers, the absence of direct engagement—such as interviews—with burnt-out mothers may have limited the depth of our insights. Third, our focus on English-language posts excludes cultural nuances and variations in how burnout is experienced and expressed by mothers across different linguistic and cultural contexts. Finally, our analysis of community responses was limited to first-level comments, potentially overlooking important communication dynamics that emerge in deeper comment threads and back-and-forth exchanges beyond the initial replies. Future work should address these limitations by incorporating cross-platform comparisons, conducting interviews or surveys to complement computational findings, performing multilingual analyses, and examining deeper reply chains to capture richer interaction dynamics.

\section{Conclusion}
\label{section:conclusion}
This work sheds light on the deeply personal and complex experiences of maternal burnout as voiced in anonymous online spaces. By leveraging Reddit data, we identified and analyzed self-disclosures from mothers experiencing burnout, revealing a wide spectrum of emotional, logistical, and relational challenges. Through a combination of machine learning techniques and qualitative analysis, we documented not only the struggles these mothers face—ranging from identity conflicts and self-doubt to unmet needs for support—but also the community responses they receive. Notably, emotional and informational support often emerged through shared experiences, empathy, and advice. Our findings highlight the pivotal role digital platforms play as support systems for mothers, potentially because offline avenues feel inaccessible due to stigma or isolation. We conclude by offering design recommendations for online platforms like Reddit to better support vulnerable populations such as burnt out mothers.

\section{Acknowledgments}

We would like to thank Ivan Maykov for his assistance in the annotation of the data used in Study 1. 

\printbibliography

\appendix

\section{Additional Details about the Methods of Our Work}

\subsection{Study 1: Data Collection \& Motherhood Detection Models}

\subsubsection{Distribution of Subreddits in the Initial Collection}
\label{appendix:methods:study1:subreddit_list}

In section \ref{section:study1:method:data_collection} we described our data collection and cleaning strategy. These steps lead to the a dataset of 28K posts (the distributions of which are displayed in Table \ref{table:list_of_subreddits}). The most repeated subreddits within this dataset are as follows: 

The top 10 parental subreddits within our dataset are as follows: \textit{r/breakingmom} (1,459), \textit{r/workingmoms} (721), \textit{r/beyondthebump} (596), \textit{r/stepparents} (362), \textit{r/daddit} (330), \textit{r/toddlers} (270), \textit{r/breastfeeding} (153), \textit{r/parentsofmultiples} (76), \textit{r/homeschool} (65), and \textit{r/regretfulparents} (60). We can see that subreddits dedicated to the discussion of issues faced by mothers include the most number of posts within our dataset. \textit{daddit}, which is among the top 10, is described as ``a sub for dads helping dads". As a result, all posts from this subreddit are removed in the final step when we characterize posts discussing mothers experiences of burnout. Another subreddit that is not in the top 10 but is eventually removed is \textit{predaddit}. Removing these two subreddits results in a dataset of 3,836 submissions from parental subreddits.

The top 10 mental health subreddits within our dataset are as follows: \textit{r/offmychest} (6,402), \textit{r/depression} (5,850), \textit{r/autism} (4,406), \textit{r/mentalhealth} (3,835), \textit{r/depression\_help} (460), \textit{r/lonely} (445), \textit{r/therapy} (344), \textit{r/socialanxiety} (285), \textit{r/mentalillness} (272), and \textit{r/selfharm} (224). While some of these subreddits have names associated with mental health issues other than burnout, we elect to keep posts from all these subreddits as the authors of all these posts discuss burnout to some extent within their posts (as they all use at least one burnout keyword). 

\subsubsection{Hyperparameter Tuning of Our Deep Learning Motherhood Detection Models} 
\label{appendix:methods:study1:dl_hyperparams}

We perform a grid search over the following hyperparameters for our model aiming to predict authorship by mothers:

\begin{itemize}
    \item $model \in \{$\textit{distilbert-base-uncased}~\cite{Sanh2019DistilBERTAD}, \textit{microsoft/deberta-v3-base}~\cite{he2021debertav3, he2021deberta}$\}$
    \item $max\_length \in \{128, 512\}$
    \item $weight\_decay \in \{0.01, 0.001\}$
    \item $learning\_rate \in \{1e^{-5}, 2e^{-5}, 3e^{-5}, 1e^{-4}\}$
    \item $per\_device\_train\_batch\_size \in \{8, 16\}$
    \item $num\_train\_epochs \in \{2, 4, 6\}$
\end{itemize}

Maximum length values are selected based on values tested in prior work~\cite{quijano2021grid, yang2019xlnet}. Posts longer than the maximum length are truncated to fit into the assigned limit. \textit{deberta} has been shown to be an improvement over \textit{roberta}~\cite{liu2019roberta} which was shown to perform well in social computing tasks.

\subsubsection{Error Analysis of Our Best Motherhood Detection Models} 
\label{appendix:methods:study:error_analysis}

The performance of our motherhood detection models are displayed in Table \ref{table:best_performing_models}. Our best performing deep learning model was a fine-tuned \textit{deberta-v3-base} model, and the best in-context learning model was a few-shot prompt with 2 randomly selected exemplars. The error analysis of these models is presented in this section. 

For our best performing in-context model, we see the number of false negatives is much lower than false positive counts. That indicates that the model is likely to classify a post as written by a mother, even if it is not. Our best-performing model configuration of 2 randomly chosen few-shot exemplars, produces 6 false negative instances. On analyzing these instances we see a pattern of the model's inability to recognize and classify posts that talk about being a mother indirectly, through experiences that would be unique to a mother, such as \textit{giving birth}, or \textit{wanting advice on breastfeeding}. 
In these posts, while being a mother is not directly referenced, the author's mention of these experiences as their own indicates that the author was a mother. Additionally, since our data is sourced from 'Reddit', the model is unable to understand jargons such as 'FTM'\textit{(first time mother)} or 'HCBM' \textit{(High Conflict Bio Mom)}.

We also analyze false positives produced by our best few-shot learning model. We find that none of these posts include a disclosure of gender either explicitly or implicitly. There is, however, a mention of life partners or relatives, usually 'husband' or 'in-laws' in the post. As annotators, since this mention is accompanied by the lack of gender direction, we don't assume the gender of the poster. However, the model might correlate due to it's implicit bias of a traditional family set-up. Furthermore, we also see that these posts are often talking about housework or house chores, with a smaller number (11\%) talking about school work. The model hence assigns a female role to the poster and classifies it as a mother. 2 posts in this stack of false negatives, call for advice from the other mothers on the subreddit by using a phrase such as: \textit{Hey Mommas}, the lack of gender and mention of this word, probably causes the model to assume that the poster is also a mother. Performing similar error analysis on our deep-learning model, the pattern for false classification are similar. However, the overall performance of the model is better. 

\subsection{Study 2: Understanding Content of Posts and Comments}

\subsubsection{Social Support: Converting 3 Classes to Binary Labels}
\label{appendix:methods:study2:social_support_converting}

To convert the 1-3 labels to 0/1 and find the label that corresponded to a minimum bar for emotional and informational support, a member of our research team (an undergraduate computer science student with prior experience in qualitative annotation) performed a manual evaluation of 60 comments. To select these comments, 10 random samples of each label (1, 2, 3) were selected for both emotional and informational support. The annotator then reviewed these 60 comments, determining if they were supportive or not (0/1). Comments labeled 2 reached the minimum threshold of emotional or informational support. As a result, all comments labeled 1 for a supportive category were converted to a label of 0, for no support, and all comments labeled >= 2 for emotional or informational support were converted to 1, demonstrating support in that category.

\subsubsection{Deep Learning Models to Predict Support}
\label{appendix:methods:study2:social_support_models}

We perform a grid search over the following hyperparameters for our model aiming to predict whether a post is supportive. A separate model is trained for each type of support (i.e., one model for emotional support and one model for informational support): 

\begin{itemize}
    \item $model \in \{$\textit{distilbert-base-uncased}~\cite{Sanh2019DistilBERTAD}, \textit{microsoft/deberta-v3-base}~\cite{he2021debertav3, he2021deberta}$\}$
    \item $weight\_decay \in \{0.001\}$
    \item $learning\_rate \in \{2e^{-5}, 3e^{-5}, 4e^{-5}\}$
    \item $per\_device\_train\_batch\_size \in \{4, 8\}$ for DeBerta, and $\{8, 16, 32\}$ for Distilbert
    \item $num\_train\_epochs \in \{2, 4, 6\}$
\end{itemize}

Table \ref{table:support_model_performance} displays the hyperparameters of the best models for detecting either type of support. 

\subsubsection{Error Analysis}
False negatives produced by the models indicate that they had difficulty identifying posts that communicated support less explicitly. For example, ``sounds like my experiences on dating websites. A rollecoaster of highs and downs until you reach a mental breakdown" was given a 0, despite relating to the original poster. The emotional support of this comment comes from the commenter's expression of their similar experiences, despite not explicitly using language supportive of the original poster.

Conversely, false positives arose from uses of mental health-related words such as "depression" or "anxiety" being used in a context that is not supportive. If a commenter was criticizing a poster's understanding of depression, that might be labeled as emotional support. 
Comments the model incorrectly labeled as informationally supportive describe emotional difficulties, but come from an angle of emotional support, rather than giving factual information. For example, "we feel guilty b/c we want so badly to have control in a situation, that by its own definition, is a complete loss of power" relates to the original poster emotionally, but is not actually giving concrete information about dealing with feelings of guilt, which might include attending therapy or seeking to alleviate the cause of guilt.
the best models (displayed in Table \ref{table:support_model_performance}) were deployed on the set of collected comments.

\end{document}